\PassOptionsToPackage{dvipsnames,table}{xcolor}
\documentclass[acmsmall,screen]{acmart}

\setcopyright{rightsretained}
\acmPrice{}
\acmDOI{10.1145/3571202}
\acmYear{2023}
\copyrightyear{2023}
\acmSubmissionID{popl23main-p72-p}
\acmJournal{PACMPL}
\acmVolume{7}
\acmNumber{POPL}
\acmArticle{9}
\acmMonth{1}

\usepackage{booktabs}   
\usepackage{subcaption} 

\usepackage{listings}
\usepackage{lstcoq}
\usepackage{lstcaml}
\usepackage{lstspecir}
\usepackage{graphicx}
\usepackage{tikz}
\usepackage{tikz-cd}
\usepackage{url}
\usepackage{xspace}
\usepackage{amsmath}
\usepackage{syntaxdefs}
\usepackage[ligature, inference]{semantic}
\usetikzlibrary{positioning}

\usepackage{wrapfig}

\lstdefinelanguage
   [x64]{Assembler}     
   [x86masm]{Assembler} 
   {morekeywords={CDQE,CQO,CMPSQ,CMPXCHG16B,JRCXZ,LODSQ,MOVSXD, %
                  POPFQ,PUSHFQ,SCASQ,STOSQ,IRETQ,RDTSCP,SWAPGS, %
                  rax,rdx,rcx,rbx,rsi,rdi,rsp,rbp, %
                  r8,r8d,r8w,r8b,r9,r9d,r9w,r9b, %
                  r10,r10d,r10w,r10b,r11,r11d,r11w,r11b, %
                  r12,r12d,r12w,r12b,r13,r13d,r13w,r13b, %
                  r14,r14d,r14w,r14b,r15,r15d,r15w,r15b}} 

\lstdefinestyle{generic}{ %
  xleftmargin=2\parindent,
  captionpos=b,
  basicstyle=\small\ttfamily,
  commentstyle=\itshape\color{MidnightBlue},
  identifierstyle=\color{black},
  keywordstyle=\bfseries,
}
\definecolor{mygreen}{RGB}{30,192,30}

\definecolor{myblue}{RGB}{30,30,192}

\definecolor{myred}{RGB}{192,30,30}
\newcommand{\outline}[1]{{\footnotesize\color{myred}{}}}

\newcommand*\figref[1]{Figure~\ref{fig:#1}}
\newcommand{\sectref}[1]{Section~\ref{sec:#1}}

\newcommand{\eg}{\emph{e.g.},\xspace}
\newcommand{\ie}{\emph{i.e.},\xspace}

\newcommand{\JITname}{FM-JIT}
\newcommand{\JIT}{{\JITname}\xspace}
\newcommand{\JITplain}{{\JITname}\xspace}
\newcommand{\IRname}{CoreIR}

\newcommand{\IRplain}{\IRname\xspace}

\newcommand{\coqin}[1]{\lstinline{#1}}
\newcommand{\specin}[1]{\lstinline[language=specir]{#1}}

\def\E0{\(\varepsilon\)}

\def\js{\ensuremath{\mathit{js}}}
\def\ms{\ensuremath{\mathit{ms}}}
\newcommand{\ftos}{\ensuremath{\text{\specin{free\_to\_state}}}\xspace}
\newcommand{\jitstep}{\ensuremath{\text{\specin{jit\_step}}}\xspace}
\newcommand{\sok}{\ensuremath{\text{\specin{SOK}}}\xspace}

\newcommand*\myxrightarrow[2]{\xrightarrow{#1} {\!\!}^{#2}\,}
\newcommand{\step}[3]{#1 \myxrightarrow{#3}{} #2}
\newcommand*\measure[1]{{m(#1)}}

\def\expr{\ensuremath{\mathit{ e}}}
\def\reg{\ensuremath{r}}

\newcommand{\funid}{\ensuremath{f}}

\def\js{\ensuremath{\mathit{js}}}

\def\nop{{\bf Nop}}

\def\call{{\bf Call}}
\def\return{{\bf Return}}
\def\cond{{\bf Cond}}

\def\print{{\bf Print}}

\def\assume{{\bf Assume}}

\mathlig{->}{\rightarrow}
\mathlig{<-}{\leftarrow}
\mathlig{!}{~!~}
\mathlig{=}{~=~}
\mathlig{++}{\text{\small++}}
\mathlig{?}{~\downarrow~}
\mathlig{/}{~\Downarrow~}

\newcommand\V{\ensuremath{\mathit{V}}}

\newcommand\fla{\ensuremath{\mathit{l}}}

\begin{document}

\title[]{Formally Verified Native Code Generation in an Effectful JIT \\ 
{\small {\it \bf or}: Turning the CompCert Backend into a Formally Verified JIT Compiler}}  


\author{Aurèle Barrière}
\orcid{0000-0002-7297-2170}
\affiliation{%
  \institution{Univ Rennes, Inria, CNRS, IRISA}
  \city{Rennes}
  \country{France}
}
\email{aurele.barriere@irisa.fr}

\author{Sandrine Blazy}
\orcid{0000-0002-0189-0223}          
\affiliation{%
  \institution{Univ Rennes, Inria, CNRS, IRISA}
  \city{Rennes}
  \country{France}
}
\email{sandrine.blazy@irisa.fr}

\author{David Pichardie}
\orcid{0000-0002-2504-1760}
\affiliation{%
  \institution{Meta}
  \city{Paris}
  \country{France}
}

\begin{abstract}
Modern Just-in-Time compilers (or JITs) typically interleave several mechanisms to execute a program.
For faster startup times and to observe the initial behavior of an execution, interpretation can be initially used.
But after a while, JITs dynamically produce native code for parts of the program they execute often.
Although some time is spent compiling dynamically, this mechanism makes for much faster times for the remaining of the program execution.
Such compilers are complex pieces of software with various components, and greatly rely on a precise interplay between the different languages being executed, including on-stack-replacement.
Traditional static compilers like CompCert have been mechanized in proof assistants,
but JITs have been scarcely formalized so far, partly due to their impure nature and their numerous components.
This work presents a model JIT with dynamic generation of native code, implemented and formally verified in Coq.
Although some parts of a JIT cannot be written in Coq, we propose a proof methodology to delimit, specify
and reason on the impure effects of a JIT.
We argue that the daunting task of formally verifying a complete JIT should draw on existing proofs of native code generation.
To this end, our work successfully reuses CompCert and its correctness proofs during dynamic compilation.
Finally, our prototype can be extracted and executed.
\end{abstract}



\maketitle

\section{Introduction}

\outline{
\begin{itemize}
\item JITs are useful.
\item What are JITs? List components. Jits are effectful. Jits directly install machine bytes for the generated code.
\item Jits are not yet formally verified.
\item List Contributions.
\end{itemize}}

Formally verified compilers are compilers that come with a machine-checked proof of correctness; it establishes that
the compiler does not introduce bugs during compilation.
Such compilers are programmed using a proof assistant, but can also be run independently.
For instance, a compiler written in the purely functional Gallina programming language of Coq can be automatically extracted to an equivalent OCaml program. 
If one trusts the extraction mechanism, the properties about the Coq code transfer to the extracted OCaml code. 
While standard (namely ahead-of-time) compilation has been the target of many verification works, such as CompCert~\cite{compcert,CACM:compcert,kastner:hal-01643290,leroy:hal-01238879},
Vellvm~\cite{vellvm,ZhaoNMZ13}, Crellvm~\cite{KangKSLPSKCCHY18}, and
CakeML~\cite{cakeml,TanMKFON16,OwensNKMT17,LoowKTMNAF19}
just-in-time compilation poses new challenges that have yet to be verified in a proof assistant.

Instead of translating a whole program once for all, then executing its output, just-in-time compilers (later referred as JIT) present an alternative approach: they start with interpretation, but parts of the program are translated at the last possible moment during execution.
This is often useful to execute dynamic languages such as Python and JavaScript~\cite{v8, pypy}, but also other languages such as Java~\cite{hotspot}, using information known only at run-time.

In essence, a JIT interleaves the execution of a program with its optimization. Optimizations are either standard (static) optimizations or JIT-specific dynamic optimizations. 
In practice, the dynamic optimization part of modern JITs often consists in dynamically generating native code for efficiency, for some parts of the program (typically functions) likely to be executed a lot. When executing a program with a JIT, the execution of high-level non-optimized code is then interleaved with the execution of native code that has been dynamically generated.
With JIT compilation, the program evolves during its execution. The program being executed at some point in time depends on what dynamic optimizations have been made before and cannot be known before execution.

\begin{figure}
\begin{center}
\begin{tikzpicture}[%
        every node/.style={rectangle,minimum size=0pt, minimum width=3cm},
        shorten >=2pt,
        node distance=0.3cm, >=latex,
        align=center
      ]
      \node [] (monitor) [] {1. Monitor};
      \node [draw] (prof) [below=of monitor, yshift=0.3cm] {Profiler};
      \node [] (opt) [below=of prof] {3. Optimization};
      \node [draw] (spec) [below=of opt, yshift=0.3cm] {Speculation};
      \node [draw] (backend) [below=of spec] {Backend\\ Compiler};
      \node [draw] (install) [below=of backend, fill=white!70!red] {\textit{Code installation}};
      \node [] (ir) [left=of opt] {2. IR Execution};
      \node [] (hl) [below=of ir, fill=white!70!red,yshift=-0.15cm] {\phantom{Sp}};
      \node [draw] (interp) [below=of ir, yshift=0.3cm] {Interpreter\\\textit{Stack manipulation}};
      \node [] (native) [right=of opt] {4. Native Execution};
      \node [draw] (run) [below=of native, yshift=0.3cm, fill=white!70!red] {Run Native\\\textit{Stack manipulation}};
      \path [draw] (prof) edge[->] node {} (ir);
      \path [draw] (prof) edge[->] node {} (opt);
      \path [draw] (prof) edge[->] node {} (native);
      \path [draw] (spec) edge[->] node {} (backend);
      \path [draw] (backend) edge[->] node {} (install);
      \node [draw] (pure) [right=of monitor, xshift=2cm, minimum width=3cm] {\footnotesize Pure and Terminating};
      \node [draw] (impure) [below=of pure, yshift=0.3cm, fill=white!70!red, minimum width=3cm] {\footnotesize Effectful Primitives};
    \end{tikzpicture}
\end{center}
\caption{Key components of a JIT with native code generation.}
\label{fig:archi}
\end{figure}
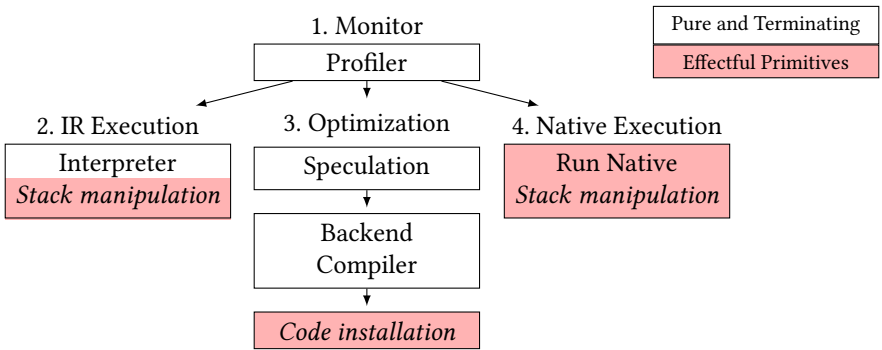

A JIT is much more complex than a standard compiler; it needs to orchestrate the interplay of several key components illustrated in \figref{archi}, including a standard compiler (the box called backend compiler).
To simplify the figure, we consider that programs are represented at an intermediate level (called IR) where optimizations and interpretation are performed, even if
some JITs rather interpret bytecode.
First, a monitor (component 1.) is in charge of gathering information and compilation hints about the program execution using a profiler, and regularly suggests functions to be optimized. This monitoring step is a simple computation that calls the relevant JIT component among three possible ones: either interpretation of the program at the IR level if there is nothing to optimize (2.) or at the native level if some function has been optimized (4.) or program optimization (3.).

A typical optimization of modern JITs is dynamic speculation.
Given some likely assumption suggested by the monitor, it allows the JIT to specialize some functions of the program at the only cost of dynamically checking that the assumption holds.
For instance, some JITs can speculate on values or types of variables in a function, and create a version of that function specialized to the assumed behavior. This is especially useful for dynamic languages.
Other optimizations such as those found in a standard backend-compiler may be triggered as well. They are followed by a code installation pass that inserts the machine bytes produced by the compiler in an executable code section of memory. When calling such a compiled function, the JIT can then jump to where it was installed, which can lead to significant speedups compared to IR interpretation. 
Moreover, JITs with speculative optimizations must include a deoptimization mechanism to switch from native code execution to the interpretation of the original function if some speculative assumption does not hold. 
This may happen in the middle of the execution of a function and requires performing \emph{on-stack replacement}: replacing the current native stackframe with a stackframe of the interpreter.

Our goal is to formally verify such a JIT in Coq, but we face several JIT-specific challenges.
Such challenges do not occur in existing verified compilers such as CompCert, that can be written as a pure and terminating Gallina function. 
First, some JIT components simply cannot be written directly in Coq, for instance the installation of bytes in memory or calling native functions.
Second, a JIT interleaves the execution of two languages, the IR and the native code.
Both share some data structures (\ie a stack and a heap).
Stack manipulation is made particularly difficult with on-stack replacement as one needs to be able to synthesize an interpreter stackframe.
There is little hope of defining mutable, global Coq data-structures that can also be accessed and modified by native code.

Our solution is to define a small set of primitives for everything that cannot be written in Coq in a JIT (\ie the pink boxes of \figref{archi}).
This includes the shared data-structures and their operations, but also installing and running machine code.
These simple primitives are not implemented in Coq, but can be specified using Coq functions on an abstract model of the JIT memory.
We use a variation of free monads, a pure functional encoding of effectful programs~\cite{datatypesalacarte}, to represent in Coq a program such as the JIT with unimplemented primitives.
Primitives can be called from both native code and components defined in Coq, allowing the interoperability that JITs intrinsically need.
We define custom calling conventions using these primitives.

For instance, consider a program being executed by our JIT containing a function \specin{F} calling a function \specin{G}, as seen on Figure~\ref{fig:ex_synchro}.
At some point during the execution of the program, the JIT may have already compiled \specin{G} to x86, as seen on the right of this Figure.
When interpreting \specin{F}, our interpreter sees the call to another function and uses a primitive to save its current environment to the JIT execution stack.
The intepreter then returns to the monitor, asking to call function \specin{G}.
The monitor then uses primitives to push the call argument to the stack, then to load and execute the x86 code corresponding to \specin{G}.
This x86 code has been generated such that it starts by using a primitive (\coqin{_Pop}) to get the call argument.
After doing the computation, it then uses another primitive (\coqin{_Push}) to push the return value of function \specin{G} to the JIT execution stack.
Finally, it returns to the monitor.
To execute the remaining of function \specin{F}, the monitor uses primitives to get the result and the top interpreter stackframe, then calls the interpreter again.

\begin{figure}
     \centering
     \begin{subfigure}[t]{0.25\textwidth}
         \centering
\begin{lstlisting}[language=specir, basicstyle=\footnotesize]
Function F (): (IR)
  x <- 1
  y <- Call G(x)
  Return y
  \end{lstlisting}
     \end{subfigure}
     \hfill
     \begin{subfigure}[t]{0.25\textwidth}
         \centering
\begin{lstlisting}[language=specir, basicstyle=\footnotesize]
Function G (a): (IR)
  b <- a+1
  Return b
  \end{lstlisting}
     \end{subfigure}
     \hfill
     \begin{subfigure}[t]{0.4\textwidth}
         \centering
  \begin{lstlisting}[language={[x64]Assembler}, basicstyle=\linespread{0.8}\footnotesize]
$Function G: (x86,simplified)
  call	_Pop
  leal	1(%eax), %edi
  call	_Push
  ret
  \end{lstlisting}
     \end{subfigure}
        \caption{Two functions of a program, where \specin{G} has been compiled to x86 using our custom calling conventions.}
        \label{fig:ex_synchro}
\end{figure}

Primitive specifications allow us to reason on the behavior of an effectful JIT with multiple languages and shared data-structures, as if the JIT was entirely programmed in Coq.
Specifically we prove that, using an abstract model of the JIT memory and an abstract specification of the primitives, executing a program with the JIT outputs the same behavior than specified by the program semantics.
The result, which is the subject of this paper, is a Coq JIT called \JIT (Free-Monadic JIT) that comes with a correctness proof, but can also be extracted and completed with primitive implementations to be executable.
If one trusts this implementation to match its specification, the Coq correctness property extends to the executable JIT.
To the best of our knowledge, \JIT is the first formally verified and mechanized JIT including 1) a standard optimizing compiler backend to dynamically produce native code,
and 2) the interoperability of native code and interpretation, with support for on-stack-replacement.
This represents a substantial step toward the verification of realistic modern JIT compilers, some of the most complex execution engines.
\JIT uses the CompCert backend to dynamically generate native code corresponding to a function.

Unless noted otherwise, all results presented in this paper have been mechanically verified
using the Coq proof assistant~\cite{Coq}. The complete development, including mechanized proofs, is
available as an artifact. 
Specifically, we claim the following novel contributions:
\begin{description}
\item[JIT design:] We introduce \JIT, a Coq JIT design with dynamic native code generation and execution.
Its dynamic compilation provides support for speculative instructions, an advanced feature of modern JITs.
This design clearly specifies each JIT component and their interplays, which are seldom described in the literature.
\item[New proof techniques:]         
      Using a variation of free monads, we develop a proof methodology for formally verifiable and executable programs with impure and non-terminating components, with a minimal trusted code base. 
      It relies on refinement to switch between different primitive specifications for more modular proofs. We apply this methodology to \JIT. 
\item[Proof reuse:]
      As JITs reuse static compilation techniques to generate native code, we argue that formally verified JITs should reuse formally verified static compilers proofs to alleviate the proof burden of such complex pieces of software.
      We demonstrate that it is possible to reuse both CompCert proof techniques (namely backward simulations) and CompCert correctness proofs in \JIT, without any modification of existing proofs.
\item[JIT correctness:]
      We prove correct that any behavior of \JITplain according to the specifications of all its impure components, is a behavior of its input program.
The proof composes the complex correctness proofs of all its components, including its backend compiler.
 \item[Runnable JIT:]
      Combining Coq extraction to OCaml and a C implementation of the primitives, we obtain a runnable JIT that dynamically generates and executes actual native code.
As expected, we observe speedups compared to interpretation alone.
\end{description}

Our work has limitations that we state here.
JITs with speculation typically insert their assumptions and specialize their functions during execution.
The insertion of dynamic speculation is out of scope of this work, and \JITplain does not include the full speculation component of \figref{archi}.
This orthogonal problem has been investigated in~\cite{corejit}.
However, we show that we can provably compile speculative instructions by allowing our input programs to be already specialized.
The simple primitive implementations used for the runnable JIT are also not proved correct, but can be audited manually and compared to their Coq specification (see Section~\ref{subsec:tcb}).
Just like CompCert, the FM-JIT correctness proof stops at the assembly level when compiling functions. We then trust an assembler to produce equivalent machine code.

This paper is organized as follows. 
First, \sectref{compcert} recalls the required background on the CompCert compiler. Then, \sectref{overview} gives an overview of \JITplain.
\sectref{freemonad} introduces our Free Monad methodology to represent and reason in Coq about an effectful program such as \JITplain. 
\sectref{proof} details our proof techniques and their application to the correctness of \JITplain. 
\sectref{implem} presents some perspectives about \JITplain as an executable JIT.
Related work is discussed in \sectref{sota}, followed by conclusions.

\outline{
Verification challenges
\begin{itemize}
\item specify each JIT component and their interplays
\item handle dynamic optimizations (done in POPL'21)
\item need to reason on a program that can not be written in Coq (need to call x86 code from a Gallina function and to modify the memory stack)
\item reuse well-known proofs of static compilers 
\item extract an executable JIT from our verified compiler, with competitive performances (wrt unverified JITs), efficiency of the generated code (requires an impure stack)
\end{itemize}
}

\section{Background on the CompCert Verified C Compiler}
\label{sec:compcert}

CompCert~\cite{CACM:compcert} is the first commercially available
optimizing compiler that is formally verified. It consists of a frontend from C to RTL (3-address code), and a backend from RTL to native code. 
CompCert targets several architectures but we focus on x86-64 assembly.
This section introduces the
ingredients of the correctness proof of its passes, 
further detailed in~\cite{Leroy-backend}. 

In CompCert, a small-step semantics defines an execution relation between
semantic states and associates to each program the set of its possible
behaviors (termination, divergence and going-wrong), including its trace of observable I/O events
(\eg calls to external library functions). 
Diverging executions observe finite or infinite traces.
A generic notion of transition semantics is defined; it consists
of a type of program states (where some are initial and others final for a terminating execution) and a step relation $\rightarrow$ over these states.
CompCert defines several transition relations (\eg star and plus transitive
closures) from the generic $\rightarrow$. 
Given a program \coqin{P} written in a language defined by its semantics \lstinline{sem}, we write 
\lstinline{program_behaves (sem p) beh} to mean that the execution of \coqin{P} according to \coqin{sem} observes a behavior \coqin{beh}.

The correctness theorem is stated as follows:
if CompCert produces code \coqin{C} from source \coqin{S}, 
then every observable behavior \coqin{bc} of \coqin{C} (\coqin{program_behaves (sem C) bc}) is a possible behavior \coqin{bs} of \coqin{S}
 (\coqin{program_behaves (sem S) bs}):
either \coqin{bc} is \coqin{bs}, or \coqin{bc} improves \coqin{bs} (\coqin{behavior\_improves bs bc}), meaning that the finite trace of events observed before \coqin{C} went wrong 
is a prefix of the trace observed during the execution of \coqin{S}.
As a compiler, CompCert is
 decomposed into several passes, and the correctness theorem
results from the correctness of each compiler pass, and so does the correctness of its backend.  
The standard technique
to prove the correctness of a pass is to prove a backward
simulation (\ie every behavior of a transformed program \coqin{C} is a behavior of
the source \coqin{S}).
It is often hard to prove a backward simulation; for passes that preserve nondeterminism, it
is easier to reason on forward simulations (\ie every behavior
of \coqin{S} is also a behavior of \coqin{C}). A backward simulation from \coqin{C} to \coqin{S} can
be constructed from a forward simulation from \coqin{S} to \coqin{C} when \coqin{C} is
deterministic. 

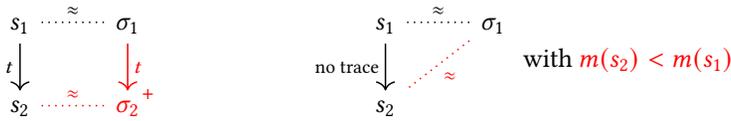
\begin{figure}
\begin{tikzcd}
  s_1 \arrow[r, dash, dotted, "\approx"] \arrow[d, "t"'] & \sigma_1 \arrow[d, red, "t"]\\
  s_2 \arrow[r, dash, dotted, red, "\approx", "+" xshift=1cm]& \textcolor{red}{\sigma_2}
\end{tikzcd}
\hspace{52pt}
\begin{tikzcd}
  s_1 \arrow[r, dash, dotted, "\approx"] \arrow[d, "\text{no trace}"'] & \sigma_1 \arrow[ld, dash, red, dotted, "\approx"]\\
  s_2
\end{tikzcd}
with \textcolor{red}{\(\measure{s_2} < \measure{s_1}\)}
\caption{Forward-simulation diagram with measure.
  Black lines are hypotheses, red lines are conclusions. On the left of each diagram are the source program and its current state $s_1$; the target program and its current state $\sigma_1$ are on the right. Vertical lines represent steps and horizontal lines the matching relation $\approx$.}
  \label{fig:diagram}
\end{figure}


Last, the most general forward-simulation diagram is defined as follows. The correctness proof of a compiler pass from
language $L_1$ to language $L_2$ relies on a forward simulation
diagram shown in \figref{diagram} and expressed in the following
theorem. Given a program $P_1$ and its transformed program $P_2$, each
transition step in $P_1$ with trace $t$ must correspond to transitions
in $P_2$ with the same trace $t$ and preserve as an invariant a
relation $\approx$ between states of $P_1$ and $P_2$. In order to
handle diverging execution steps and rule out the infinite stuttering
problem (that may happen when infinitely many consecutive steps in
$P_1$ are simulated by no step at all in $P_2$), the theorem uses a
measure over the states of language $L_1$ that strictly decreases in
cases where stuttering could occur. It is generically noted
\(\measure\cdot\) and is specific to each compiler pass. 
In CompCert's parlance, this diagram is denoted by
\specin{forward_simulation (sem1 P1) (sem2 P2)}, where \specin{semi} defines
the semantics of $L_i$.

\section{Overview of \JITplain and its Correctness Theorem}\label{sec:overview}

This section presents the salient features of \JITplain: its architecture, source language, impure primitives, and its main correctness theorem.
While minimal, it precisely captures the essence of the dynamic generation of native code and its execution in a JIT.

\subsection{The Architecture of \JITplain}
\outline{Detail each component in OUR design. Here we explain that we do not insert speculation.}

\begin{figure}
  \centering
\begin{tikzpicture}[%
        every node/.style={rectangle,minimum size=0pt, minimum width=15pt},
        shorten >=2pt,
        node distance=0.5cm, >=latex
      ]
      \node [draw] (prof) [inner sep=5pt] {MONITOR};
      \node [] (prof') [below=0.5mm of prof] {profiler};
      \node [draw] (opt) [below left=of prof', inner sep=2pt] {OPTIMIZATION};
      \node [draw] (exe) [below=1.5cm of prof'] {DISPATCH};
      \node [draw] (exeir) [below left=of exe, yshift=-0.5cm, xshift=-2cm, inner sep=5pt] {IR EXECUTION};
      \node [draw] (exenat) [right=of exeir, inner sep=5pt, xshift=3cm] {NATIVE EXECUTION};
      \node [] (call) [above=0.5mm of exenat, xshift=1.4cm] {native call};
      \node [] (int) [above=0.5mm of exeir, xshift=-2cm] {interpreter};
      \path [draw] (prof') edge[->]  node {} (opt)
      (opt) edge[->, left] node {} (exe)
      (prof') edge[->]  node {} (exe)
      (exe) edge[->]  node {} (exeir)
      (exe) edge[->]  node {} (exenat)
      (exeir) edge[->, loop left, left] node {} (exeir)
      (exeir) edge[->, bend left=60, left, align=center] node {return, call,\\deoptimization\phantom{a}} (prof.west)
      (exenat) edge[->, dashed, bend right=60, right, align=center] node {return, call,\\deoptimization} (prof.east);
    \end{tikzpicture}
    \caption{A JIT architecture as a state machine. Dashed line represents multiple steps (see Section~\ref{subsec:nonatomic}).}
    \label{fig:monitor}
\end{figure}
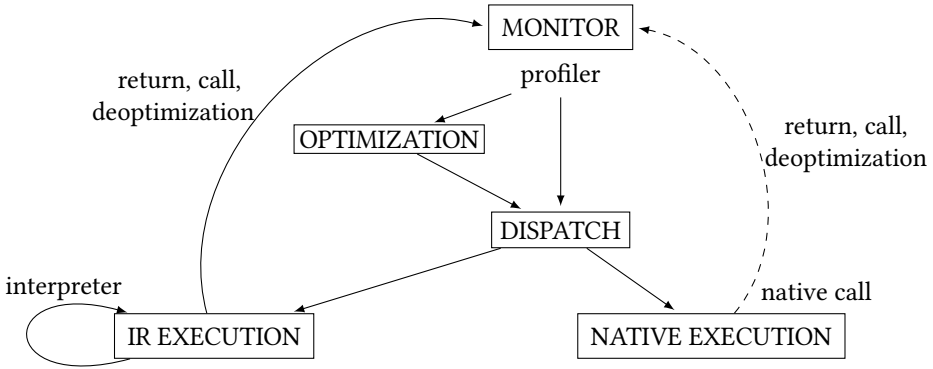

A JIT resembles an interaction loop in that it executes or optimizes the code until the program finishes.
Figure~\ref{fig:monitor} shows the state machine that describes how the JIT monitor alternates between its components, from IR execution, to optimization and native execution.
\JITplain is defined with a function representing the transitions of this state machine and calling the relevant JIT component.
Each transition can be represented as a terminating Gallina function (possibly using primitives).
However, the execution of native code may be non-terminating if the JIT compiled a diverging function.
This is why one transition of the JIT is represented in a dashed line in \figref{monitor}, meaning that it corresponds to a sequence of elementary transitions.
We first ignore that non-atomic transition, then explain how it is defined and specified in Section~\ref{subsec:nonatomic}.

The profiler corresponds to a collection of heuristics that inspect the current execution and suggest optimizations.
The profiling heuristics themselves are out of scope of our verification work, as they are often highly empirical.
Our formalization takes as input any of these heuristics.
The benefits of clearly separating these from the rest of \JITplain is that our correctness results do not depend on the implementation of such heuristics, which
 should only have an impact on performance, but not on correctness.

An optimization is made of three steps. First, the IR function to compile is translated into 
RTL (see Section~\ref{subsec:rtlgen}).
Second, the CompCert backend is called on the translated RTL to produce some x86 code.
Third, this x86 code is installed in an executable portion of the memory.
This last part cannot be written directly in Gallina, and is part of the primitives we specify (as listed on Section~\ref{subsec:primitives}).

Whenever the profiler did not suggest any optimization, the JIT resumes the program execution.
Depending on whether or not the current function has been compiled yet, execution is dispatched to either the interpreter or some dynamically generated native code.
This execution component stays active until the execution reaches a \textit{synchronization point} (either a function call, or a function return, or a deoptimization trigger), where it returns to the monitor.
The monitor is then tasked with managing the stack. It may push arguments to the stack if the synchronization point is a function call, or even synthesize a new interpreter stackframe if the synchronization point is a deoptimization trigger.
After a deoptimization, execution will always be dispatched to IR execution, since the new current stackframe is an interpreter one. 

\subsection{The Low-level Intermediate Representation \IRname}\label{sec:ir}
\outline{Like CoreIR but simplified. 
Will be augmented with RTL and x86 dynamically.
Speculation and deoptimization still possible but has to be pre-inserted.
I do not think we should write down its semantics: say that this is standard. Later we'll see the mixed one.}

The source language that \JITplain executes is \IRname. 
Its syntax is depicted Figure~\ref{fig:coreir-syntax}.
There is an infinite number of registers \textit{r}, and each can contain a value $v$, namely a 64-bit integer.
A function has an identifier \textit{f}, an entry point \textit{l} and a list of arguments $r^*$. Its code is represented by its control-flow graph, namely a mapping from labels \textit{l} representing program points to instructions \textit{i}.
Functions can have up to two versions, an optional one where speculation has been inserted, and an original one to deoptimize to if needed.

\begin{figure}[!t]
\begin{subfigure}[b]{0.4\textwidth}
\begin{syntax}
\syntaxclass{Expressions:}
\expr  &::= & \reg~+~\reg~|~\reg~-~\reg~|~\reg~*~\reg & \\
       &\mid& \reg~=~\reg~|~\reg~<~\reg~|~\reg~\%~\reg~|~-~\reg  & \\
       &\mid& \mathit{v}~|~\reg~=~0~|~ \reg~+~\mathit{v} ~|~\reg~*~\mathit{v} & \\
\end{syntax}
\end{subfigure}
\begin{subfigure}[b]{0.5\textwidth}
\begin{syntax}
\syntaxclass{Programs:}
\V
       &::= &      l \mapsto i                  & Code\\
{\it F}
       &::= &      \{\reg^*,l, \V,{\it option~\V}\}  & Function \\
{\it P}
       &::= &      \funid \mapsto {\it F} & Program\\
\end{syntax}
\end{subfigure}
\vspace{-0.5cm}
\centering
\begin{syntax}
\syntaxclass{Instructions:}
i      &::= &    \nop~l  \,\,\, \mid \,\,\, \reg <- ~\expr ~l                & no-operation and operations\\
       &\mid&    \cond~\reg~l_t~l_f   \,\,\, \mid \,\,\,   \print~\reg~l    & branching, output value\\
       &\mid&    \reg<-\call~\,\funid~\reg^*~l  \, \,\, \mid \,\,\,   \return~\reg  & call and return\\
       &\mid&    \reg <- \textbf{MemGet}~\reg_{ad}~l   \,\,\, \mid \,\,\,    \reg_{ad} <- \textbf{MemSet}~\reg~l & memory load and store\\
       &\mid&    \assume~\reg~\funid.\fla~[(\reg <- \expr)^*]~l & speculation\\
\end{syntax}
\vskip -0.8em
\caption{Syntax of \IRplain.}\label{fig:coreir-syntax}
\end{figure}

\IRplain was introduced in~\cite{corejit} as an extension of  RTL, with some instructions for speculation, inspired by~\cite{sourir}.
The \textbf{Assume} instruction represents a speculation. To execute it, its condition (what is being speculated on) is dynamically checked. If it holds, execution continues to the next label.
Otherwise, deoptimization happens, and the JIT monitor must build an interpreter stackframe following the deoptimization metadata contained in that instruction.
For instance, \specin{Assume x F3.l2 [a}$\leftarrow$\specin{5, b}$\leftarrow$\specin{y] l4} will check if \coqin{x} is true (any value other than 0). If it is, execution jumps to \specin{l4}. Else, execution deoptimizes into the original version of function \specin{F3}, at label \specin{l2}. To reconstruct the environment of that function, we put 5 in register \coqin{a} and evaluate \coqin{y} in the current environment for the value of \coqin{b}.
This work shows that such inserted speculations can be correctly compiled down to native code by the backend of \JITplain.
A CoreIR program also gets access to a part of the memory, the heap (a fixed-size array of 64-bit integer values), using  \textbf{MemGet} and \textbf{MemSet} instructions.

\IRplain is  simplified compared to~\cite{corejit}.
First, the \textbf{Anchor} instruction, useful to insert speculations, has been removed since speculation insertion is out-of-scope of \JITplain.
Second, the \textbf{Assume} instruction is simplified to only allow the synthesis of a single stackframe at once when deoptimizing. Synthesizing multiple stackframes has to be done when inlining functions, but \JITplain does not perform inlining and instead compiles functions one at a time.
Finally, the operations have been modified to better resemble  RTL.
The semantics of \IRplain is close to the RTL semantics and the semantics of the \textbf{Assume} instruction is defined in ~\cite{corejit}.

\subsection{The Minimal Set of Impure Primitives of \JITplain}
\label{subsec:primitives}
\outline{In the next section we'll see how to write the JIT without these parts.
First we define the list of everything we're not going to write in Coq.
List them all.
Without CloseSF. Maybe not OpenSF.}

This section describes the parts of \JITplain that cannot be written in Coq.
Since their implementation has to be trusted, we keep the smallest possible interface of impure primitives that a JIT with native code generation requires.
\JITplain uses the following impure primitives:
\begin{itemize}
\item \coqin{HeapGet x} to get the heap value at address \coqin{x}.
\item \coqin{HeapSet x y} to set \coqin{x} in the heap at address \coqin{y}. 
\item \coqin{Push x}, to push a value \coqin{x} on the stack.
\item \coqin{Pop}, to pop a value from the top of the stack.
\item \coqin{Push_IRSF sf} to add an interpreter stackframe \coqin{sf} to the stack.
\item \coqin{OpenSF} to get the top stackframe of the stack.
\item \coqin{Install_Code asm} to install some native code \coqin{asm} generated by the optimizer.
\item \coqin{Load_Code fun_id} to load some installed native code for function \coqin{fun_id}.
\item \coqin{Check_installed fun_id} to check if some function \coqin{fun_id} has been compiled.
\item \coqin{Print x} to print some value \coqin{x}.
\end{itemize}

In static compilation, one simply generates code that modifies the memory, while in a JIT the memory is a data-structure directly modified during execution.
\coqin{HeapSet} and \coqin{HeapGet} are used to modify and access the heap.
These primitives are called by the IR interpreter when interpreting \textbf{MemSet} and \textbf{MemGet} instructions, but also by the native code of a function that originally contained these instructions.
Section~\ref{subsec:free_monads} details how primitives are called from Coq components and an example of our generated native code is shown later on Figure~\ref{fig:ex_x86}.

The execution stack contains both interpreter stackframes and stackframes for the native code.
Stackframes for the native code consist in 64-bit integers pushed on top of the stack.
The native code we generate uses custom calling convention, involving \coqin{Pop} and \coqin{Push} to add or get integers from the stack, for instance to get function arguments or push return values. Examples of these custom conventions are shown in Section~\ref{subsec:rtlgen}.
When deoptimizing, the native code also pushes its deoptimization metadata to the stack using \coqin{Push}.
The monitor uses both primitives when calling a native function (pushing arguments), or when returning from a native call (popping the return value), or after native-code deoptimization (popping the deoptimization metadata before creating the corresponding stackframe).
But the stack also contains interpreter stackframes, that can be pushed by the interpreter with \coqin{Push_IRSF} when an interpreted function calls another.
Technically, we could design our interpreter such that it uses stackframes made of 64-bit integers as well, and then use the \coqin{Push} primitive for both native and interpreter stackframes.
Such an interpreter would be more difficult to write and prove correct than our current version using these Coq records, which closely follows the \IRplain semantics.
Having a dedicated primitive also shows that our design is not tied to a particular interpreter implementation.
After a function return, the monitor uses \coqin{OpenSF} to get the top stackframe, and dispatches execution accordingly.

Finally, \JITplain must also install the native code it generated.
The optimizer uses \coqin{Install_Code} to allocate some space in the memory, write machine bytes and make that memory executable.
This also includes calling an assembler to produce machine code from the assembly code generated by the JIT.
The addresses of these allocations are stored, and \coqin{Load_Code} can then be used when running native code to return the address of the machine code corresponding to a function identifier.
\coqin{Check_Installed} is used by the monitor to decide if it should call the interpreter or the native code for function calls.
Note that tracking which function has been installed could be done purely in Coq, removing this last primitive from the list.
We include this primitive anyway to define a comprehensive interface of everything a JIT needs to do with its executable memory.
Note that the three parts of the JIT memory (execution stack, heap and executable codes) are disjoint.

As JITs differ from static compilers in their need to do effectful impure computations, this list already sheds some light on the way a formally verified JIT should be designed.
The impure effects of \JITplain  are restrained to that small list of impure primitives.
Executing native code is yet another impure computation done by the JIT that is handled differently and explained in Section~\ref{subsec:nonatomic}.
All other things done by \JITplain can be written directly in the pure programming language of Coq.

\subsection{The Correctness Theorem of \JITplain}
\label{subsec:theorem}
\outline{Just the theorem, not the explanation for the proof.
We cannot yet define the JIT semantics, we'll see that in the next section.
It might require an explanation of CompCert simulations if we want both the final theorem and the backward.
Introduce refinement: we will prove it for any implementation that satisfy the spec (2 Coq theorems).}

CompCert simulations state that any behavior of a compiled program matches a behavior of its source program.
While the program of a JIT dynamically evolves during execution, one can still prove a similar correctness theorem if we compare the semantics of the input program to some small-step semantics (called \coqin{jit_sem}) describing the behavior of \JITplain executing a program.
In short, every step  consists in a transition of the state machine of \figref{monitor}.
Many of these transitions are pure Coq functions that can be used directly to define small steps.
Other use some of the impure primitives listed in Section~\ref{subsec:primitives}. These cannot be written as Coq functions, but can be specified with Coq monadic functions.
\coqin{jit_sem} is defined in Section~\ref{subsec:monad_spec}, where the specification of the primitives is used to define the remaining transitions.
Last, we reuse the CompCert simulation framework (see \sectref{compcert}) to prove \JITplain correct by proving the following theorems (see details in Section~\ref{subsec:jit_correct}), where \coqin{prim_spec} is a specification
of the impure primitives (see Section~\ref{subsec:monad_spec}), and \coqin{CoreIR_sem} is the small-step semantics of the \IRplain language.

\begin{lstlisting}[language=Coq]
Theorem jit_correctness_simulation:
  forall p, backward_simulation (CoreIR_sem p) (jit_sem p prim_spec).
Theorem jit_correctness:
  forall p beh, program_behaves (jit_sem p prim_spec) beh ->
    exists beh', program_behaves (CoreIR_sem p) beh' /\ behavior_improves beh' beh.
\end{lstlisting}\label{th:main}

The above theorem states that for any source program \coqin{p}, if the JIT behaves in some behavior 
\lstinline{beh}, then \lstinline{beh} improves a behavior of the original program semantics.
Behavior improvement means that if \coqin{p} goes wrong, \JITplain is allowed to avoid going wrong and can produce any behavior \coqin{beh} after that point.
A corollary states that if \coqin{p} does not go wrong, then any JIT behavior is exactly a behavior of \coqin{p}.
This theorem strongly resembles the one from CompCert, only replacing the semantics of the compiled program with the semantics \coqin{jit_sem} of \JITplain.

Note that \JITplain is not allowed to go wrong if \coqin{p} does not.
In particular, when the compilation of a function fails (because some analysis did not converge), \JITplain simply cancels the optimization and keeps interpreting safely.
Similarly, some primitives may fail during execution of \JITplain, but we prove that they go wrong only when the source program goes wrong.
For instance, heap accesses can fail if the index is out of bounds. 
We thus know that \JITplain will only execute a failing heap access if the source program did so.

\section{Specifying \JITplain with a Monadic Encoding}
\label{sec:freemonad}
\outline{The free monad methodology to represent a program with effects like the JIT.
We write exactly what we want to extract to OCaml, and leave the rest out.}

A static compiler like CompCert is written as a pure and terminating program.
Hence, it is a prime candidate for a traditional extraction workflow: one can write a compiler as a Coq function, then one can extract that function to an equivalent executable OCaml function.
However, JITs are impure and effectful programs.
Yet, these impure parts (\eg calling native code or interacting with global data-structures) are not the only things a JIT does.
A JIT must also translate code (with its backend compiler), interpret code, and orchestrate its components (with its monitor).
All these remaining parts, making up for most of a JIT's work, can be written in a pure functional language and can be formally verified.
In this section, we are tasked with the challenge of verifying in Coq a program that can only be partially written in Coq.

Our solution is to design a formalism inspired by free monads to write incomplete Coq programs, that
may contain pure computations, but may also contain calls to some unimplemented impure primitives.
In short, these incomplete programs contain all of the pure parts of the program, and can be written in Coq.
In fact, free monads have been used for years in pure functional languages to represent incomplete programs~\cite{datatypesalacarte}.
Incomplete programs are not executable yet, as they lack some implementations.
\JITplain can be written as an incomplete program, using the list of primitives of Section~\ref{subsec:primitives}, and also writing every pure computation of a JIT: most of the monitor, dynamic compiler, and interpreter.

Obviously, this is not enough on its own, we want a JIT that is both verified and executable.
For the verification of such an incomplete program, we can fill its holes with specifications for the primitives.
In Section~\ref{subsec:state_monads}, we present state and errors monads, which provide a good formalism to encode in a pure functional language all of our primitive specifications.
For the JIT, we can write pure Coq monadic functions working on an abstract model of the JIT memory to specify each primitive.
The free monad used to write \JITplain is defined in Section~\ref{subsec:free_monads}.
Next, Section~\ref{subsec:monad_spec} shows that an incomplete program can be completed with such primitive specifications.
The result is a Coq function that specifies the behavior of the complete program.
We use that Coq function to define the small-step semantics of \JITplain that we can then prove correct.

To get an impure executable JIT from such an incomplete program, we must move away from Coq.
We extract it to an incomplete OCaml program.
In OCaml, where effects are possible and C functions can even be called, we define impure implementations of each primitive, working on actual shared data-structures and really calling native code.
In Section~\ref{subsec:impure_prims}, we show that both the incomplete program and the primitive implementations can be composed together to get an executable impure program.

Using free monads not only has the advantage of clearly identifying and specifying the various effectful components of a JIT, it also allows us to switch between different specifications of our data-structures.
This is the \textit{refinement} methodology of Section~\ref{subsec:refinement}, facilitating the proofs of a multi-language JIT.
Moreover, the meta-theory needed to reason about our free monad implementation is very lightweight (a few hundreds of Coq lines).
In particular, we only ever use free monads to write terminating computations, which eliminates the need for coinductive reasoning. We show in Section~\ref{subsec:nonatomic} that our monads can be used in conjunction with the small-step semantics framework of CompCert to reason about possibly infinite behaviors.

There are two main drives behind our chosen formalism.
First, the meta-theory used to reason about effectful programs should be as simple and lightweight as possible, to not hinder already difficult proofs.
Second, our formalism should be entirely compatible with CompCert so that we can reuse both its correctness proof and its simulation framework, which
already handles proof composition and diverging executions. 
In the end, this free monad formalism could be used for any verification work trying to extend the CompCert simulation framework to effectful programs.

\subsection{An Existing Solution for Specifying Effects: State and Error Monads}
\label{subsec:state_monads}
\outline{Coq definition. Bind, ret.
Already in CompCert.
Good enough to specify, but not to actually implement.}

\begin{figure}
\centering
\begin{subfigure}[t]{0.4\textwidth}
\begin{lstlisting}[language=Coq,basicstyle=\footnotesize]
Inductive sres (state A:Type) : Type :=
| SError : errmsg -> sres state A
| SOK : A -> state -> sres state A.

Definition smon (state A:Type) : Type :=
  state -> sres state A.
\end{lstlisting}
\end{subfigure}
\begin{subfigure}[t]{0.5\textwidth}
\begin{lstlisting}[language=Coq, basicstyle=\footnotesize]
Definition sret state A (x:A) : smon state A :=
  fun (s:state) => SOK x s.
Definition sbind state A B (f: smon state A)
               (g:A -> smon state B) : smon state B :=
  fun (s:state) => match (f s) with
    | SError msg => SError msg
    | SOK a s' => g a s'
    end.
\end{lstlisting}
\end{subfigure}
\caption{Defining the Coq state and error monad.}
\label{fig:statemonad}
\end{figure}

A standard way to encode global effects in pure functional languages is to use \textit{state monads},
an abstraction of computations modifying a global state.
In \JITplain, we use a variation of the state monad that also includes errors: our primitives may modify a global state, but also fail (for instance, when trying to pop an empty stack).
That state and error monad definition can also be found in CompCert.
The CompCert definition of state monads that we reuse is on the left of Figure~\ref{fig:statemonad}.
Intuitively, a state monad of type \coqin{smon A} represents computations that either return a value of type \coqin{A} and possibly change the global state, or fail.
The \coqin{smon} type is parameterized by a type \coqin{state} representing global states, which contain a model of the stack, heap, and executable codes.
The type \coqin{sres} represents the possible return values of the monads.
Such state monads are executable Coq functions, taking as argument an initial global state, and returning its return value as well as the new global state (or an error).

Like all monads, state monads come with helpful constructors \coqin{sret} and \coqin{sbind} to build complex monadic computations, as defined on the right of Figure~\ref{fig:statemonad}.
The  constructor \coqin{sbind} sequentially chains together monadic computations, constructing entire programs that may have global effects.
For instance, executing \coqin{sbind f g} in some global state \coqin{s} first executes \coqin{f} on \coqin{s}.
If that computation succeeded and returned a value \coqin{a} and modified the global state to \coqin{s'}, then we execute \coqin{g} on \coqin{a} and \coqin{s'}.

We could write \JITplain as a state monadic computation whose state contains a model of the stack, the heap and the generated native codes.
However, extracting such a JIT to OCaml would only produce a pure OCaml program, because the Coq extraction only targets a pure subset of OCaml.
With such an approach, there is no hope to get an executable JIT that installs and calls actual native code.
In the end, state monads are a great formalism to specify primitives that modify a global state, but this is not enough to implement an effectful executable JIT.

\subsection{Our Solution to Write \JITplain in Coq: Free Monads}
\label{subsec:free_monads}
\outline{Coq definition. Bind, ret. 
Example of a component written as a Free Monad (optimizer).}

Since the primitives cannot be written in Coq, \JITplain is an incomplete Coq program.
Free monads are a convenient formalism to write incomplete programs, namely programs with some holes left to represent the primitives that have not yet been implemented.
Free monads provide a DSL to write such programs given a list of primitives that they may use.
Such incomplete programs will either be completed with specifications of the primitives (Section~\ref{subsec:monad_spec}), or with impure implementations of the primitives (Section~\ref{subsec:impure_prims}).
First, we inductively define the primitives our free monad definition uses as on the left of Figure~\ref{fig:freemonad}. 
We see that one possible primitive is \coqin{Prim_Push}, taking an \coqin{int} as argument and not returning any value.

\begin{figure}
\centering
\begin{subfigure}[t]{0.5\textwidth}
\begin{lstlisting}[language=Coq, basicstyle=\footnotesize]
Inductive primitive: Type -> Type :=
| Prim_Push: int -> primitive unit
| Prim_Pop: primitive int
| Prim_HeapSet: int -> int -> primitive unit
| Prim_HeapGet: int -> primitive int
...
\end{lstlisting}
\end{subfigure}
\begin{subfigure}[t]{0.45\textwidth}
\begin{lstlisting}[language=Coq, basicstyle=\footnotesize]
Inductive free (T :Type) : Type :=
| pure (x : T) : free T
| impure R (prim : primitive R)
     (cont : R -> free T) : free T
| ferror (e : errmsg) : free T.
\end{lstlisting}
\end{subfigure}
\caption{Definition in Coq of free monads.}
\label{fig:freemonad}
\end{figure}

Effectful computations are defined on the right of Figure~\ref{fig:freemonad}.
A term of type \coqin{free T} is called a free computation; it encodes a function that computes some value of type \coqin{T}, possibly using some primitives.
The computation is either a pure computation, not using any primitive, an error, or an impure computation.
Such an impure computation calls an unimplemented primitive \coqin{prim}, and then a continuation \coqin{cont} encodes the rest of that computation (possibly using other primitives), given the return value of the primitive.
This simple type is quite easy to manipulate.
For instance, one can write LTac tactics that automatically check that some free computation only uses some primitives and
this helps us prove that the interpreter does not install any new native code.

\begin{figure}
\begin{subfigure}[t]{0.5\textwidth}
\begin{lstlisting}[language=Coq, basicstyle=\footnotesize]
Fixpoint fbind X Y
   (f: free X) (g: X -> free Y) : free Y :=
  match f with
  | pure x => g x
  | impure R prim cont =>
     impure prim (fun x => fbind (cont x) g)
  | ferror e => ferror e
  end.
Definition fret X (x:X) : free X := pure x.
Definition fprim R (p:primitive R) : free R :=
  impure p fret.
\end{lstlisting}
\end{subfigure}
\begin{subfigure}[t]{0.4\textwidth}
\begin{lstlisting}[language=Coq, basicstyle=\footnotesize]
Notation "'do' X <- A ; B" :=
  (fbind A (fun X => B)).

Definition optimizer (f:function) : free unit :=
  do f_rtl <- fret (IRtoRTL f);
  (* using CompCert backend *)
  do f_x86 <- fret (backend f_rtl);
  (* impure computation *)
  fprim (Prim_Install_Code f_x86).
\end{lstlisting}
\end{subfigure}
\caption{Free monadic constructors and using them to write \JITplain.}
\label{fig:free_cons}
\end{figure}

Free monads are monads, and as such come with the monadic constructors on the left of Figure~\ref{fig:free_cons}.
Binding impure free computations entails adding primitives to the continuation.
These constructors allow us to define complex free computations, such as all the components of \JITplain.
For instance, the optimizer step is detailed on the right of Figure~\ref{fig:free_cons}.
This code compiles and installs some function. First we call \coqin{IRtoRTL} which generates RTL code. This is a pure computation, using no primitive (hence the \coqin{fret}).
We then call the CompCert backend that generates some equivalent x86 code.
Finally, we call a primitive, \coqin{Prim_Install_Code}, that installs the generated code in an executable portion of the memory.
While this looks like a simple program, \coqin{backend} is a pure function containing all of the CompCert backend, from RTL to x86.
Here, the free monad encoding is used to orchestrate complex pure transformations with calls to the impure interface.
To be executable, such a computation requires an implementation for \coqin{Prim_Install_Code}.

The entire JIT can be represented this way.
As explained in Section~\ref{sec:overview}, defining \JITplain consists in defining the transition of a state machine.
We can then describe the JIT as a function \coqin{jit_step} with the type \lstinline[language=Ocaml]{jit_state -> free (jit_state * trace)}, where \coqin{jit_state} is the type of states of the state machine, as on Figure~\ref{fig:monitor}, but also contains the CoreIR functions and the data updated by the profiler.

\subsection{Monadic Specifications and Semantics of Free Monads}
\label{subsec:monad_spec}
\outline{How to specify each primitive. 1 example of a spec.
The exec function: transforming monads into Coq functions that can define a small-step semantics (for now, the non-atomicity is not an issue).}

\JITplain is an incomplete Coq program, written using free monads and lacking some implementation for the primitives it uses.
In order to derive its semantics \coqin{jit_sem}, we specify each primitive using a state and error monads (with the definitions of Section~\ref{subsec:state_monads}).
Since our primitives work on global data-structures and may fail, the state and error monad is adequate for specifying them.
We define in \figref{spec} a \textit{monadic specification} as a record containing an initial global state and a state-monad computation for each primitive.
It is parameterized by the type of global states, containing a model of the stack, the heap, and the executable installed code of \JITplain.
Moreover, we  define \coqin{get_prim} to access the corresponding specification of a primitive with its arguments.
An example of specification is given in \figref{impl_and_spec} for the \coqin{heap_get} primitive.

\begin{figure}
\begin{subfigure}[t]{0.45\textwidth}
\begin{lstlisting}[language=Coq, basicstyle=\footnotesize]
Record monad_spec (mstate:Type): Type :=
  mk_mon_spec { 
    init_state : mstate;
    prim_push: int -> smon mstate unit; 
    prim_pop: smon mstate int;
    prim_heapset: int -> int -> smon state unit;
    prim_heapget: int -> smon state int
    ... }.
\end{lstlisting}
\end{subfigure}
\begin{subfigure}[t]{0.45\textwidth}
\begin{lstlisting}[language=Coq, basicstyle=\footnotesize]
Definition get_prim R S (p:primitive R)
              (i:monad_spec S) : smon S R :=
  match p with
  | Prim_Push x => (prim_push i) x
  | Prim_Pop => (prim_pop i)
  | Prim_HeapSet x y => (prim_heapset i) x y
  | Prim_HeapGet x => (prim_heapget i) x
  ...
\end{lstlisting}
\end{subfigure}
\caption{Coq monadic specifications of representative primitives.}
\label{fig:spec}
\end{figure}

Furthermore, to fill the holes of incomplete programs, a function called \coqin{free_to_state} in \figref{freetostate} transforms any free monad computation \coqin{f} into a state monad computation.
It simply replaces recursively any call to a primitive \coqin{prim} by its specification.
It uses \coqin{get_prim} to get the primitive specification, and the continuation of an impure computation is bound to the result using the state-monad bind.
\begin{figure}
\begin{lstlisting}[language=Coq, basicstyle=\footnotesize]
Fixpoint free_to_state (A S:Type) (f:free A) (i:monad_spec S): smon S A := match f with
  | pure a => sret a
  | ferror e => fun (s => SError e)
  | impure R prim cont => sbind (get_prim prim i) (fun r:R => free_to_state (cont r) i)
end.
\end{lstlisting}
\caption{Turning free computations into state and error computations.}
\label{fig:freetostate}
\end{figure}

Now that free computations can be completed with primitive specifications, we can define the small-step semantics of the entire JIT.
State and error monadic computations are executable Coq functions, so one can execute the one corresponding to the JIT transitions.
The semantics states of the JIT execution contain both a \coqin{jit_state} $(\js_1)$, the data that can be written and manipulated in Coq (a state of Figure~\ref{fig:monitor} also including the \IRplain functions), and a state of the state-monadic specification \coqin{mstate} $(\ms1)$, a model of the data structures that we cannot write in Coq. One simply goes from one of such state to another according to the execution of the state-monad computation given by completing the JIT free transitions \coqin{jit_step}. 
The single small-step semantic rule is defined in Figure~\ref{fig:jit_sem}, where $i$ is a monadic specification, and $t$ is the observed trace. In the figure, we omit the types \coqin{A} and \coqin{S}, implemented respectively with \coqin{jit_state * trace} and the \coqin{mstate} type of $i$.

\begin{figure}
\begin{gather*}
\inference[\coqin{jit_sem}]{
    \ftos \, \, (\jitstep \,\, \js_1) \, \, i \, \, \ms_1 = \sok \, \, (\js_2, t) \, \, \ms_2
}{ 
\step {(\js_1,\ms_1)} {(\js_2,\ms_2)} t
}
\end{gather*}
\caption{The JIT small-step semantic rule.}
\label{fig:jit_sem}
\end{figure}


\subsection{An Impure Implementation for  \JITplain}
\label{subsec:impure_prims}
\outline{Example of C primitives.
The OCaml loop.}

Once completed it with monadic specifications, we extract \JITplain to OCaml and complete it with impure and effectful implementations of the primitives.
The result is an effectful OCaml JIT that can be executed.
We write C functions for interacting with our global data-structures.
For instance, \figref{impl_and_spec} shows the C implementation for the heap access primitive.
On the right is its monadic specification used in the JIT semantics.
While the C function accesses some global array \texttt{jit\_heap}, the specification accesses a map contained in its monadic state \coqin{s}. Both the global array and the monadic state are unchanged and the primitive fails for out-of-bound accesses.

\begin{figure}
     \centering
     \begin{subfigure}[t]{0.45\textwidth}
     \begin{lstlisting}[language=C, basicstyle=\footnotesize]
int64_t heap_get (int64_t x){
  assert (x < HEAP_SIZE);
  int64_t val = jit_heap[x];
  return val; }
     \end{lstlisting}
     \end{subfigure}
     \begin{subfigure}[t]{0.45\textwidth}
     \begin{lstlisting}[language=Coq, basicstyle=\footnotesize]
Definition heap_get (x:int) : smon int :=
  fun s => if (Int.lt x heap_size) then
      SOK (PMap.get (pos_of_int x) (heap s)) s
      else SError ("MemGet out of memory range").
     \end{lstlisting}
     \end{subfigure}
   \caption{A C primitive implementation and its Coq monadic specification, that reuses CompCert libraries \coqin{Int} (to compare 64-bit integers) and \coqin{PMap}. }
   \label{fig:impl_and_spec}
\end{figure}

\begin{figure}
\begin{lstlisting}[language=Ocaml, basicstyle=\footnotesize]
let rec free_interpreter (f: A free) : A = match f with
  | Coq_pure (a) -> a
  | Coq_ferror (e) -> print_error e; failwith "Free monad error"
  | Coq_impure (prim, cont) -> let x = exec_prim prim in free_interpreter (cont x)   end.  
\end{lstlisting}
\caption{Executing free computations in OCaml.}
\label{fig:free_interpreter}
\end{figure}

Next, \figref{free_interpreter} shows an interpreter of free monad computations in OCaml, where \coqin{exec_prim} executes the C function corresponding to the primitive.
It directly interprets the incomplete program itself, without resorting to state monads and \coqin{free_to_state}.
Running our free computation \coqin{jit_step} through this \coqin{free_interpreter} until the program execution finishes, we get an executable JIT in OCaml that calls effectful primitives.

\subsection{Facilitating the Correctness Proofs with Refinement }
\label{subsec:refinement}
\outline{Get closer to the impure implementation.
An advantage of the free monad methodology. We do not have to describe the exact refinement we prove.
Show the refinement theorem: if we prove JIT correct on the naive spec, we can get the same one on another.}

In our approach there is a small list of JIT-specific functions to manually audit: our primitives implementations must match their specifications.
One could then try to write specifications that closely match what the impure implementation is doing, but in practice that monadic specification can be hard to reason with in the JIT correctness proof.
In particular, for a fast access to the stack using only \coqin{Pop} and \coqin{Push} in the generated native code, one would like the execution stack to be a simple array of integers.
However, when writing our proof invariants, it would be simpler if these integers were structured in several lists, each list corresponding to a specific stackframe.
Moreover, in the final executable implementation, the execution stack is split in two parts: one that holds the interpreter stackframes, and one that holds the integer stackframes.
Some simulation invariants (\eg in Section~\ref{subsec:rtlgen}) are however easier to write if there is a single stack containing both interpreter and native stackframes. Then, proving the compilation of a function correct simply requires substituting its future stackframes instead of moving them from one structure to the other.

We argue that an advantage of using free monads is the ability to switch between different monadic specifications.
One can then define a monadic specification that is close to the actual primitive implementations (called \coqin{prim_spec}), and another \textit{reference} monadic specification (or \coqin{ref_spec}), with which proofs are easier to conduct.
In this section, we show that once we prove that \coqin{prim_spec} \textit{refines} \coqin{ref_spec}, then the correctness theorems about the JIT semantics using the reference implementation can be propagated to the JIT semantics using \coqin{prim_spec}.
We use this technique to switch between two specifications of the execution stack: an unstructured stack of integers, close to the C implementation, and a reference one where the stack is structured for easier stack invariants. 
This refinement methodology takes advantage of free monads to modularly separate the correctness arguments for stack manipulation and stack representation.

The benefit is a more modular reasoning: first we prove correct the JIT using \coqin{ref_spec}, then we prove the refinement. 
So, as composing CompCert simulations facilitates modular proofs, 
instead of re-developing new proof techniques for modularity, we define our refinement relation so that we can reuse the simulation composition technique of CompCert.
More precisely, a monadic specification $i$ \textit{refines} another $j$ if there exists a relation $\simeq$ between monad states of $i$ and monad states of $j$ (written $i \simeq j$ ) such that for each primitive $p$,
\begin{align*}
\forall s_i~s_j~s_i'~\mathit{args}~r, & \quad s_i \simeq s_j \quad \wedge \quad p_i ~\mathit{args}~ s_i = \sok ~r ~s_i' \rightarrow \\
\exists s_j', & \quad s_i' \simeq s_j' \quad \wedge \quad p_j ~\mathit{args}~ s_j = \sok ~r ~s_j'
\end{align*}
where $p_k$ is the state monad for $p$ in the monadic specification $k$. 
This definition purposely resembles the forward-simulation definition of CompCert, but relating primitive executions instead of small-step semantics.
We then prove the following theorem, using the refinement relation to build a simulation invariant.

\begin{minipage}{\linewidth}
\begin{lstlisting}[language=Coq]
Theorem refinement:
  forall (prog_state istate jstate:Type) (prog:prog_state -> free (prog_state * trace))
    (i:monad_spec istate) (j:monad_spec jstate),
    refines i j -> forward_simulation (jit_sem prog i) (jit_sem prog j).
\end{lstlisting}
\end{minipage}
As the JIT behavior is deterministic, that forward simulation is used to construct a backward simulation.
Using that theorem on \JITplain and the two monadic specifications discussed above, we prove the main correctness theorem of Section~\ref{subsec:theorem} with the reference implementation, and then propagate that correctness theorem to the semantics using \coqin{prim_spec}, without additional proof effort.
We see the refinement theorem being used in Section~\ref{subsec:jit_correct}.
Proving the refinement (\coqin{prim_spec} $\simeq$ \coqin{ref_spec}) is straightforward.
In practice, while proving that an unstructured stack corresponds to a structured stack (a Coq list of stackframes containing themselves lists of integers) is not a difficult thing, its justification should not hinder the proofs of every program transformation.

\subsection{Non-Atomicity of Transitions: Small-step Semantics of x86 to the Rescue}
\label{subsec:nonatomic}
\outline{We still have an issue with the call to native code; it may not terminate.
However, we have a small-step semantics for the native code.
How we specify with 3 free monads, and how we define the small-step semantics.}

As in the \coqin{optimizer} example of Figure~\ref{fig:free_cons}, almost every transition of the JIT state-machine is written as a terminating function of type \coqin{free T} for some return type \coqin{T}.
However, more than effectful transitions, a JIT compiler may also include non-terminating transitions, as the compiled program may not terminate.
The dashed transition of Figure~\ref{fig:monitor} represents a possibly diverging computation: the JIT may have compiled and executed a non-terminating function.
This is not an issue with the IR interpreter which can be defined with some fuel (\ie an integer limiting its maximum number of steps) and return regularly to the monitor only to be called again.
But the dynamically generated native code may be stuck in a loop without any return, call or deoptimization, and the fuel technique cannot be used to represent such an infinite execution.
This possibly diverging transition cannot be specified like other primitives, with a terminating free computation.

One solution could be to extend our free monad to a coinductive structure to represent and reason about possibly non-terminating effectful programs, like in~\cite{itrees}.
However, this would require coinductive reasoning on such transitions.
On the other hand, the CompCert simulations are already equipped to reason about non-terminating executions, as long as such executions are defined by small-step semantics.
Moreover, x86 semantics in CompCert are already specified with executable small-step semantics, looping a function that computes the next semantic state.\footnote{More precisely, most of x86 semantics are defined with a function in CompCert. External calls are specified with an inductive non-executable parameter. However, in the subset of x86 that our JIT generates, the only such calls are calls to our JIT primitives that we can specify as usual with Coq state-monad functions.}
As a result, the non-atomic transition of the JIT can be decomposed into several atomic small steps of CompCert x86 semantics.
This has two advantages: not only can we avoid writing coinductive proofs in Coq and instead reuse those of CompCert, but this also facilitates the reuse of the CompCert correctness theorem, expressed in terms of these x86 semantics, without modifying it.

\begin{figure}
  \centering
\begin{tikzpicture}[%
        every node/.style={rectangle,minimum size=0pt,minimum height=4pt, inner sep=5pt},
        shorten >=2pt,
        node distance=0.3cm, >=latex, align=center
      ]
      \node [] (h) [draw] {\footnotesize DISPATCH};
      \node [] (c) [draw, right=of h, xshift=1.5cm] {\footnotesize NATIVE\\\footnotesize EXECUTION};
      \node [] (r) [draw, below=of h, yshift=-1.5cm] {\footnotesize MONITOR};
      \path [draw] (h) edge[->, bend left, below]  node {\footnotesize Atomic} (c)
      (c) edge[->,bend left=60, dashed,right]  node {\footnotesize LoadAndRun} (r.east)
      (r) edge[->, bend left, left]  node {\footnotesize Atomic} (h);
      \node [] (h') [draw, right=of c, xshift=1cm] {\footnotesize DISPATCH};
      \node [] (c') [draw, right=of h', xshift=1.5cm] {\footnotesize NATIVE \\\footnotesize EXECUTION};
      \node [] (r') [draw, below=of h', yshift=-1.5cm] {\footnotesize MONITOR};
      \node [] (i1) [draw, circle, below=of c', xshift=-0.5cm, inner sep=1pt] {\footnotesize x86};
      \node [] (i2) [draw, circle, below left=of i1, inner sep=1pt] {\footnotesize x86};
      \node [] (i3) [draw, circle, below left=of i2, inner sep=1pt] {\footnotesize x86};
      \path [draw] (h') edge[->, bend left, below]  node {\footnotesize Atomic} (c')
      (c') edge[double, ->, bend left, right] node {\footnotesize \coqin{start_}} (i1)
      (i1) edge[double, ->, below right] node {\footnotesize \coqin{step_}} (i2)
      (i2) edge[double, ->, below right] node {\footnotesize \coqin{step_}} (i3)
      (i3) edge[double, ->, bend left, below right] node {\coqin{end_}} (r'.east)
      (r') edge[->, bend left, left]  node {\footnotesize Atomic} (h');
    \end{tikzpicture}
    \caption{Giving small-step semantics to non-atomic transitions in a NASM.}
    \label{fig:unf_sem}
\end{figure}
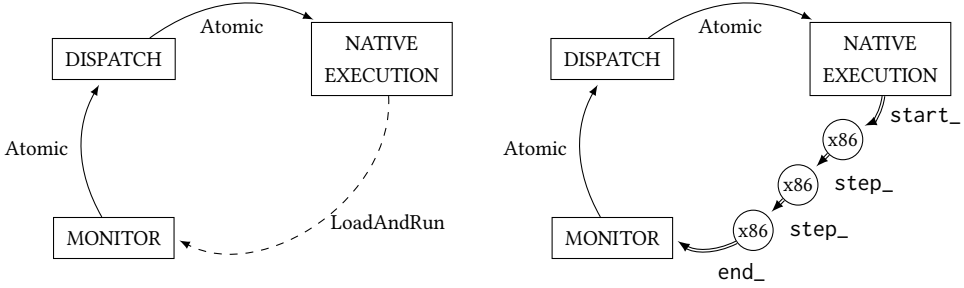

To specify such native calls, we first define \JITplain as a \textit{Non-Atomic State Machine} (or NASM), whose transitions, defined on Figure~\ref{fig:nasm}, are either atomic steps or a possibly infinite sequence of native code steps. In that definition, \coqin{Trace} is the type of observable events, and \coqin{State} is the type of state-machine states.
Next, we extend the OCaml \lstinline[language=Ocaml]{free_interpreter} of Section~\ref{subsec:impure_prims} so that it loads and calls native code when seeing a \coqin{LoadAndRun} transition.
Such a transition is only available from the \coqin{NATIVE EXECUTION} state of Figure~\ref{fig:monitor}.
Finally, to give small-step semantics to NASM, we specify the call to native code with three free computations \coqin{step_}, \coqin{start_} and \coqin{end_} as pictured on Figure~\ref{fig:unf_sem}. 
The \coqin{step_} function is the small-step transition of x86, as defined in CompCert.
We simply extend it so that every call to one of our primitives is specified with its monadic effect.
Moreover, we extend the small-step semantics definition of Figure~\ref{fig:jit_sem} with rules that execute \coqin{start_} when seeing a \coqin{LoadAndRun} transition (building an initial x86 semantic state), then loop \coqin{step_} until a final x86 semantic state is reached, and execute \coqin{end_} to get back the next JIT state.
In the case of a diverging native execution, these new JIT semantics will stay in x86 semantic states as intended, and the entire JIT behavior will be considered diverging.

\begin{figure}
\begin{lstlisting}[language=Coq, basicstyle=\footnotesize]
Inductive nasm_transition (State Trace:Type): Type :=
| Atomic: free (Trace * State) -> nasm_transition
| LoadAndRun: nasm_transition.
\end{lstlisting}
\caption{Non-Atomic State Machine definition.}
\label{fig:nasm}
\end{figure}

With these simple definitions unfolding native calls according to their CompCert semantics, one can define small-step semantics for the entire JIT even in the presence of non-atomic transitions.
And this allows us reusing the CompCert proof without modification (see Section~\ref{subsec:backend_correct}).

\section{Proving the Correctness of \JITplain}\label{sec:proof}
\outline{How we reuse CompCert and its proof to get a formally verified JIT.}
\outline{The issues with reusing CompCert: theorem does not preserve the effects on the stack and heap.
But if we compile only part of the program, we need that preservation.
So we make the generated code use JIT primitives to interact with them.
The stack and heap are external to the CompCert world.}

As JITs reuse code transformations of static compilers to generate native code,
we argue that formally verifying the backend compiler of a JIT should reuse the formal verification of static compilers.
In this section, we show that by carefully transforming our IR, we can directly reuse the proof of the CompCert backend, in order to prove the correctness of the native-code generation in \JITplain.
Intuitively, one could think that we can directly use this backend dynamically to transform some parts of the JIT program, and then because it generated ``equivalent'' code, the JIT execution semantics should not change.
However, this is not as straightforward.

First, the CompCert backend correctness theorem is established by relating the semantics of an x86 program to the semantics of a RTL one.
This means that our dynamic compilation step should be split in two passes: first \JITplain generates a piece of RTL for a given function, then it uses the CompCert backend to generate some x86 from that RTL function.
To prove these passes modularly, we need to reason about the intermediate program, where some RTL has been generated but not yet compiled.
The JIT semantics of Figure~\ref{fig:jit_sem} contains semantic states for the execution of x86 and CoreIR, but nothing for RTL since the RTL is never executed by the JIT.
To reason about our compilation step modularly, we define mixed semantics in Section~\ref{subsec:mixed} that include semantic states for CoreIR, x86 but also RTL.
These semantics are used to specify the backend of \JITplain.



Finally, the correctness theorem of CompCert states that the observable behavior of a program is preserved. 
Nothing in this theorem is related to the effects on the memory, that are indeed not preserved by the CompCert backend.
In fact, the backend goes from an abstract stack in RTL (a list of RTL stackframes), to an actual execution stack.
This is an issue for \JITplain. We want to be sure that every modification to the heap done by the RTL function will be compiled to some x86 code that also modifies the heap similarly, otherwise executing the rest of the program after that function may differ.
To avoid that issue while not modifying the code of the CompCert backend, we make the generated code interact with the stack and the heap only through external calls to 5 of the \JITplain primitives (Section~\ref{subsec:rtlgen}). This means not relying on CompCert to compile function calls, but instead generating RTL code that uses our primitives.
We then define custom calling conventions relying on our primitives that the generated code uses.
For instance, even though CompCert only compiles programs with no arguments, our solution consists in generating RTL programs that start by popping their arguments off the stack.
In the end, the native programs we generate may use the CompCert memory to spill registers, however for \JITplain, we do not use the stack and heap handled by CompCert, but rather the shared data-structures manipulated by free monads. 

\subsection{Splitting RTL Programs to Directly Reuse CompCert Proofs}
\label{subsec:rtlgen}
\outline{
2 steps: generating RTL, then generating native.
List the 5 primitives that can be accessed from native.
Schema or example of a function being compiled.}

CompCert only allows the compilation of complete programs and uses its own calling conventions.
By generating several pieces of code that interact with the stack only through our JIT primitives, we demonstrate that both CompCert and its theorem can be used for formally verified native code generation in \JITplain.
Our backend compilation process has two steps. For a given function to compile, we first generate several RTL programs. Then we compile each of them using the CompCert backend.
The verification of these two passes is discussed in Sections~\ref{subsec:callconv} and \ref{subsec:backend_correct}.

To reuse the CompCert backend with custom calling conventions, we split our \IRplain functions at function calls when generating RTL.
The only JIT primitives that can be called from that RTL code are \coqin{HeapGet}, \coqin{HeapSet}, \coqin{Push}, \coqin{Pop} and \coqin{Print}.
The first two are used to interact with the heap. Stack primitives \coqin{Pop} and \coqin{Push} are used to save the live environment, but also store return values or function arguments. 
The last one is used for any compiled function with \textbf{Print} instructions.

\begin{figure}
\begin{center}
\includegraphics[height=3cm]{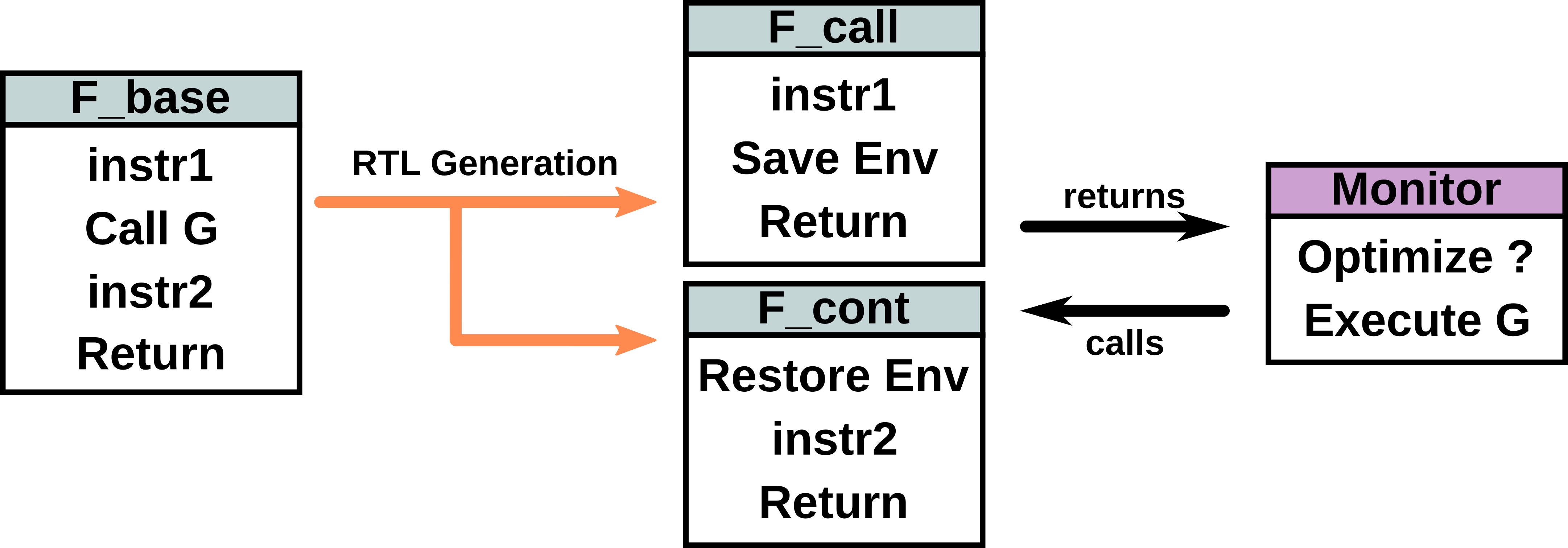}
\end{center}
\caption{Transforming \IRplain function \coqin{F_base} into two RTL functions \coqin{F_call} and \coqin{F_cont} using custom calling conventions to return to the monitor between calls.}
\label{fig:rtlgen}
\end{figure}

Figure~\ref{fig:rtlgen} shows how to split each function \lstinline{F} before compiling it.
When \JITplain decides to optimize \lstinline{F}, it first splits its original version \lstinline{F_base} in two functions: \lstinline{F_call} and \lstinline{F_cont}, its continuation after the call.
We add primitives to these functions that save and restore the environment (the live registers).
Finally, \lstinline{F_call} and \lstinline{F_cont} can be compiled with a backend that preserves the primitive calls. Each one is defined as a whole RTL program.
After optimization, \JITplain will start by calling the compiled function \lstinline{F_call}. When it encounters the call to function \lstinline{G}, the generated native code returns to the JIT monitor, which may now optimize or simply execute \lstinline{G}.
When this call returns, the monitor calls the continuation function \lstinline{F_cont}.

\begin{figure}
     \centering
     \begin{subfigure}[c]{0.3\textwidth}
         \centering
\begin{lstlisting}[language=specir, basicstyle=\footnotesize]
Function Fun1 (reg1):
  reg2 <- reg1 + 4
  reg3 <- Call Fun7(reg2)
  reg3 <- reg1 + reg3
  Return reg3  
  \end{lstlisting}
         \caption{A \IRplain function.}
         \label{fig:ex_coreir}
     \end{subfigure}
     \hfill
     \begin{subfigure}[c]{0.3\textwidth}
         \centering
\begin{lstlisting}[basicstyle=\linespread{0.8}\footnotesize]
$1() {
  x8 = "Pop"()            
  x9 = x8 + 4 (int)
  x1 = "Push" (x8)
  x1 = "Close"(1, 2)
  x1 = "Push"(x9)
  x1 = "Push"(1)
  x1 = "Push"(7)
  x7 = RETCALL
  return x7 }
  \end{lstlisting}
  \begin{lstlisting}[basicstyle=\linespread{0.8}\footnotesize]
$2() {
  x10 = "Pop"()
  x8 = "Pop"()
  x10 = x8 + x10
  x1 = "Push"(x10)
  x7 = RETRET
  return x7 }
  \end{lstlisting}
         \caption{Two Generated RTL programs.}
         \label{fig:ex_rtl}
     \end{subfigure}
     \hfill
     \begin{subfigure}[c]{0.3\textwidth}
         \centering
  \begin{lstlisting}[language={[x64]Assembler}, basicstyle=\linespread{0.8}\footnotesize]
# Generated by CompCert
$2:
leaq  32(%rsp), %rax
movq  %rax, 0(%rsp)
movq  %rbx, 8(%rsp)
call  _Pop
movq  %rax, %rbx
call  _Pop
leal  0(%eax,%ebx,1),%edi
call  _Push
movl  $RETRET, %eax
movq  8(%rsp), %rbx
addq  $24, %rsp
ret
  \end{lstlisting}
         \caption{The x86 program for the continuation.}
         \label{fig:ex_x86}
     \end{subfigure}
        \caption{Compilation of a function \coqin{Fun1} by \JITplain.}
        \label{fig:ex_compile}
\end{figure}

Figure~\ref{fig:ex_compile} shows an example of a \IRplain function being compiled.
The \coqin{Fun1} function does a computation, then calls another function \coqin{Fun7}, then does another computation and returns.
When generating RTL, because there is only one call in \coqin{Fun1}, we split the function into two RTL programs.
The first one (\coqin{$1}) starts by getting the function argument (\texttt{reg1}/\texttt{x8}) off the stack.
After an instruction for the computation, it performs a call by first saving the live register \texttt{x8} on the stack. It then closes the current stackframe by pushing the identifier of the current function and the label of the call (to identify the corresponding continuation function). \coqin{Close} is simply implemented by several calls to \coqin{Push}.
After that, we push the call arguments, the number of arguments, and the identifier of the function we want to call (\coqin{Fun7}).
Finally, we return with the constant \texttt{RETCALL}, returning to the monitor but indicating that the function wants to call another one.
The monitor may now pop the function identifier and decide to optimize or execute \coqin{Fun7}.
After the call to \coqin{Fun7}, its return value has been pushed to the stack.
The execution then follows with the second program \coqin{$2}; it
starts by getting the return value of \coqin{Fun7}, and restores the live register \texttt{x8} that was pushed earlier. Finally, it ends by returning another constant, \texttt{RETRET} to indicate to the monitor that the execution has finished.
The CompCert backend then provably preserves the calls to JIT primitives, as seen on the x86 code produced for the second RTL program of Figure~\ref{fig:ex_compile}.

During the RTL generation pass, \textbf{Assume} instructions are compiled as branches.
If the speculation holds, we proceed in the rest of the program.
Else, we push the deoptimization metadata on the stack and return with another constant \texttt{RETDEOPT}. If deoptimization occurs, we return to the monitor which will read that data from the stack and reconstruct the corresponding interpreter state.

\subsection{Mixed Semantics: Interleaving Pieces of Executions Related to Three Languages}
\label{subsec:mixed}
\outline{Show how we reuse the semantics from all 3 languages.
But also we get a monad for all effects when calling primitives.}

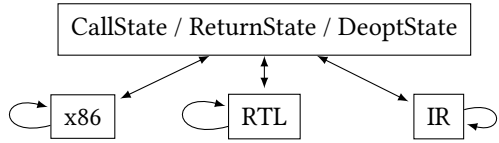
\begin{figure}
  \centering
\begin{tikzpicture}[%
        every node/.style={rectangle,minimum size=0pt,minimum height=4pt, inner sep=5pt},
        shorten >=2pt,
        node distance=0.5cm, >=latex
      ]
      \node [] (synchro) [draw] {CallState / ReturnState / DeoptState};
      \node [] (rtl) [draw, below=of synchro] {RTL};
      \node [] (x86) [draw, left=of rtl, xshift=-1cm] {x86};
      \node [] (ir) [draw, right=of rtl, xshift=1cm] {IR};
      \path [draw] (synchro) edge[<->, left]  node {} (x86)
      [draw] (synchro) edge[<->, left]  node {} (rtl)
      [draw] (synchro) edge[<->, right]  node {} (ir);
      \path [draw] (x86) edge[->, loop left] node {} (x86);
      \path [draw] (rtl) edge[->, loop left] node {} (rtl);
      \path [draw] (ir) edge[->, loop right] node {} (ir);
    \end{tikzpicture}
    \caption{Semantic states of the mixed semantics.}
    \label{fig:mixed_states}
\end{figure}

The compilation done by \JITplain is done in two passes, and its correctness proof can also be decomposed into two correctness proofs using a simulation for each pass.
This allows us to prove the splitting of functions and the use of our custom calling conventions (generating RTL) independently from the correctness of native code generation.
This modular approach to compilation correctness using simulations that are then composed together is the same strategy used by CompCert to prove each of its passes independently.
To express these two simulations, we need to define formal semantics for the intermediate JIT programs that are obtained after running our first RTL generation pass, jut like CompCert defines formal semantics for its intermediate languages.
In the case of \JITplain, such intermediate programs contain \IRname, RTL and x86 code, and in this section we present their formal semantics, called \textit{mixed semantics}.
Note that programs with pieces of RTL are never executed by the JIT, which waits until the backend compilation has entirely finished, and every piece of RTL is removed before resuming execution.
The semantic states of these mixed semantics include the semantic states of each of these three languages semantics.
To interface them, we also define three states forming a synchronization interface: CallState, ReturnState and DeoptState.
Succintly, each function call, return or deoptimization goes through shared synchronization states; this is shown on Figure~\ref{fig:mixed_states}.


Figure~\ref{fig:mixed_sem} shows representative rules of the mixed semantics.
The semantic states are pairs, of one mixed state of Figure~\ref{fig:mixed_states} and one monadic state of the reference monadic specification (containing the stack and the heap).
First, one stays in a language according to its semantics until reaching a function call, a return or a deoptimization.
For instance, in rule (step~x86), we follow the execution of the x86 semantics. On any instruction that is not an external call, the monadic state is unchanged. There is a similar rule reusing RTL semantics, (step~RTL).
However, we extend both these semantics with monadic rules when calling JIT primitives.
For instance, in rule (push~x86), if the current x86 instruction calls the primitive \coqin{Push}, we update the monadic state $ms$ with the execution of the primitive specification.
We move to the next x86 state \coqin{s'} with function \coqin{next_state}, moving to the return address.
Along with (step~x86), these rules form the modified x86 small-step semantics that we also use to specify the \coqin{step_} function of Section~\ref{subsec:nonatomic}.

When the x86 or RTL semantics reach a final state, we move to the corresponding synchronization state.
For instance in (x86~return), upon seeing the constant \coqin{RETRET}, we move to a ReturnState.
In our x86 and RTL programs, the return value is pushed on the stack, so we move to a state that does not contain the return value itself, but rather some indication (\coqin{OnStack}) that it has to be popped.
In rule (call~x86), we see one way to go from a synchronization state to a x86 state.
If we were about to call function \coqin{f} with some arguments \coqin{args}, and see that \coqin{f} had been compiled to some native program \coqin{p}, then we would push arguments to the stack and move to the initial semantic states of \coqin{p}, mimicking the behavior of the monitor and extending it to RTL executions.

Mixed semantics include other similar rules for \IRplain and RTL.
A final rule also steps from a ReturnState to a final semantic state if the stack is empty.
While CallStates and ReturnStates step to any of the three languages, DeoptStates always reconstruct an interpreter state for \IRname.
Finally, all rules depicted on Figure~\ref{fig:mixed_sem} are silent and produce an empty trace.
The only time an observable behavior is produced is when executing the primitive \coqin{Prim_Print}, either when calling it from x86 or RTL, or when interpreting a \IRplain \textbf{Print} instruction. These are the observable events preserved by the JIT execution.

\begin{figure}\small
\begin{gather*}
\inference[step x86]{
    $\coqin{s}$ ~~~->^{x86}~~~ $\coqin{s'}$ & \text{not external call}
}{
    $\coqin{(s, ms)}$ ~~~->~~~ $\coqin{(s', ms)}$
}
\quad\quad\inference[step RTL]{
    $\coqin{s}$ ~~~->^{\text{RTL}}~~~ $\coqin{s'}$ & \text{not external call}
}{
    $\coqin{(s, ms)}$ ~~~->~~~ $\coqin{(s', ms)}$
}
\\[2mm]
\inference[push x86]{
    $\coqin{find_instr(s) = Call Prim_Push [v]}$ &
    $\coqin{next_state (s, retval) = s'}$ \\
    $\coqin{free_to_state (Prim_Push v) ms = SOK retval ms'}$
}{
    $\coqin{(s, ms)}$ ~~~->~~~ $\coqin{(s', ms')}$
}
\\[2mm]
\inference[x86 return]{
    $\coqin{s}$ ~~~->^{x86}~~~ $\coqin{Final RETRET}$
}{
    $\coqin{(s, ms)}$ ~~~->~~~ $\coqin{(ReturnState OnStack, ms)}$
}
\\[2mm]
\inference[call x86]{
    $\coqin{free_to_state (Prim_Load_Code f) ms = SOK p ms}$ \\
    $\coqin{free_to_state (push_args args) ms = SOK tt ms'}$
}{
    $\coqin{(CallState f args, ms)}$ ~~~->~~~ $\coqin{(initial_state p, ms')}$
}
\end{gather*}
\caption{Some representative rules of the mixed semantics.}
\label{fig:mixed_sem}
\end{figure}

\subsection{Correctness of RTL Generation}
\label{subsec:callconv}
\outline{Theorem: backward simulation
Schema of the invariant.
It's here we prove the correctness of our Calling Conventions.
It's here that we substitute interpreter stackframes with native ones.
Talk about RTLblock?}

The first compilation pass generates several RTL programs for a given \IRplain function \coqin{F}: one program for the \coqin{F} entry, and one continuation program for each function call in \coqin{F}.
The generated code must contain the calls to JIT primitives that are going to be preserved by the CompCert backend and used by the native code.
This means that this pass stops producing interpreter stackframes but instead uses native stackframes. Proving this pass correct then means proving correct the change of calling conventions in a function.
Our proof is a forward simulation relating the mixed semantics of the current JIT program before and after transforming to RTL a given \IRplain function.

Building our invariant is crucial to proving the simulation.
There are three main cases in the invariant for the transformation of a function \texttt{F} to RTL.
First, because we are only compiling a single function, we need to relate identical semantic states when outside of that function (\emph{refl}). However, even in that case, the execution stack can differ: some interpreter stackframes for \texttt{F} may have been replaced with equivalent native stackframes containing the live registers at the time of the call.
Another possible case of the invariant (\emph{rtl}) happens when executing the new RTL function (or one of the continuation). RTL semantic states are related to \IRplain semantic states.
The two states must agree on live registers (not all registers, as only the live ones are restored after a call).
Finally, the synchronization states (Callstate, Returnstate or Deoptstate)  differ when reached from RTL or \IRname. In RTL, the arguments of such states (like call arguments, the return value or the deoptimization metadata) have been pushed to the stack instead of directly given by the interpreter. A last invariant case \emph{synchro} expresses that.

Figure~\ref{fig:rtlblock_proof} showcases an example of the invariant preservation in our simulation proof.
The execution on the left corresponds to executing the program before \texttt{F} is transformed to RTL.
Before calling the transformed function \texttt{F}, semantic states are related with the \emph{refl} invariant.
Then, we prove that the beginning of the execution of \texttt{F} in \IRplain matches its execution in RTL using the \emph{rtl} invariant, even though in RTL the arguments have to be popped first.
As we see a function call to \texttt{G} (at \coqin{l2}), the \IRplain interpreter simply builds a Callstate containing the arguments. In RTL however, we need to push that to the stack and end on a different Callstate. Both are related with the \emph{synchro} invariant.
Execution of \texttt{G} then proceeds; when it returns after several steps, the source execution simply goes back in the middle of the \IRplain \texttt{F}. On the RTL side, we show that by going into the corresponding continuation program \texttt{F.2} and popping the return value and environment off the stack, we get to semantic states matched with the \emph{rtl} invariant.

In practice, this proof is conducted in two steps: we first generate RTLblock, a language we wrote where labels can be associated to basic blocks of instructions (instead of single instructions like in RTL). We then transform that RTLblock code into RTL code, unfolding the basic blocks.
To that end, we extended the mixed semantics of Section~\ref{subsec:mixed} to also include the semantics of RTLblock.
Using RTLblock as an intermediate language between \IRplain and RTL allows us to have simpler invariants, where every \IRplain instruction is matched with a single basic block.
Both steps are proved correct with forward simulations that we compose.

\def\ghostlbl{\phantom{\specin{l2:_}}}
\begin{figure}
\begin{centering}
\begin{tikzpicture}[%
        every node/.style={rectangle, rounded corners, minimum size=4pt,minimum height=4pt, inner sep=5pt, text width=4.1cm, align=left, transform shape},
        shorten >=2pt,
        node distance=0.3cm, >=latex, scale=0.8
      ]
      \node [] (callf') [draw]
      {\textbf{Callstate} \texttt{F [Fargs]}};
      \node [] (rtlentry) [draw, below=of callf']
      {\textbf{RTL Function F:} \\
      \ghostlbl\specin{Pop Fargs}\\
      \specin{l1: instr1} \\
      \specin{l2: Push env, F, l2}\\
      \ghostlbl\specin{Push Gargs, G}\\
      \ghostlbl\specin{Return RETCALL}};
      \node [] (callg') [draw, below=of rtlentry]
      {\textbf{Callstate} \texttt{OnStack}};
      \node [] (return') [draw, below=of callg', yshift=-0.3cm]
      {\textbf{Returnstate} 32};
      \node [] (rtlcont) [draw, below=of return']
      {\textbf{RTL Continuation F.2:}\\
      \ghostlbl\specin{Pop (return value)}\\
      \ghostlbl\specin{Pop env}\\
      \specin{l3: instr3}};

      \node [] (callf) [draw, left=of callf', xshift=-2.5cm] 
      {\textbf{Callstate} \texttt{F [Fargs]}};
      \node [] (irf1) [draw, left=of rtlentry, xshift=-2.5cm]
      {\textbf{\IRplain Function F:} \\
      \specin{l1: instr1} \\
      \specin{l2: Call G (Gargs)}};
      \node [] (callg) [draw, left=of callg', xshift=-2.5cm]
      {\textbf{Callstate} \texttt{G [Gargs]}};
      \node [] (return) [draw, left=of return', xshift=-2.5cm]
      {\textbf{Returnstate} 32};
      \node [] (irf2) [draw, left=of rtlcont, xshift=-2.5cm]
      {\textbf{\IRplain Function F:} \\
      \specin{l3: instr3}};
      
      \path [draw] (callf.east) edge[-, above right]  node {\emph{refl}} (callf'.west)
      (irf1.east) edge[-, above right]  node {\emph{rtl}} (rtlentry.west)
      (callg.east) edge[-, above right]  node {\emph{synchro}} (callg'.west)
      (return.east) edge[-, above right]  node {\emph{refl}} (return'.west)
      (irf2.east) edge[-, above right]  node {\emph{rtl}} (rtlcont.west);
      \path [draw] (callf.south) edge[->, right]  node {} (irf1.north)
      (irf1.south) edge[->, right]  node {} (callg.north)
      (callg.south) edge[->, right]  node {$\ast$ \small Execution of G} (return.north)
      (return.south) edge[->, right]  node {} (irf2.north);
      \path [draw] (callf'.south) edge[->, right]  node {} (rtlentry.north)
      (rtlentry.south) edge[->, right]  node {} (callg'.north)
      (callg'.south) edge[->, right]  node {$\ast$ \small Execution of G} (return'.north)
      (return'.south) edge[->, right]  node {} (rtlcont.north);
    \end{tikzpicture}
    \end{centering}
    \captionsetup{justification=centering}
    \caption{Preservation of the invariant while transforming Function \coqin{F} to RTL.}
    \label{fig:rtlblock_proof}
\end{figure}
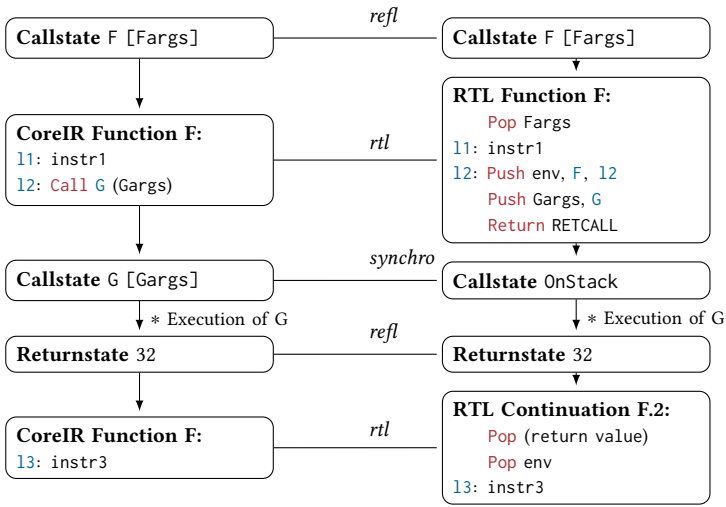

\subsection{Correctness of Native Code Generation}
\label{subsec:backend_correct}
\outline{Theorem: backward simulation
We reuse the CompCert backward simulations.
From a CompCert semantics (where the primitives are observable) to the mixed semantics (where they are silent).
Insists that these 2 sections are specific to dynamic compilation.}

\def\hhspace{\hspace{1cm}}
\newcommand{\dashdownarrow}{\raisebox{2.0ex}{\rotatebox{-90}{$\dashrightarrow$}}}
\begin{figure}
  \centering
\begin{tikzpicture}[%
        every node/.style={rectangle,minimum size=0pt,minimum height=4pt, minimum width=30pt, inner sep=5pt},
        node distance=1cm, >=latex,
        align=center
      ]
      \node [] (rtl1) [draw] {\coqin{ms}\hhspace\coqin{rtl_1}};
      \node [] (rtl2) [draw, below=of rtl1] {\coqin{ms}\hhspace\coqin{rtl_2}};
      \node [] (rtl3) [draw, below=of rtl2] {\coqin{ms'}\hhspace\coqin{rtl_3}};
      \node [] (rtl4) [draw, below=of rtl3] {\coqin{ms'}\hhspace\coqin{rtl_4}};
      \node [] (x861) [draw, right=of rtl1, xshift=1cm] {\coqin{x86_1}\hhspace\coqin{ms}};
      \node [] (x862) [draw, below=of x861] {\coqin{x86_2}\hhspace\coqin{ms'}};
      \node [] (legend) [below=of x862, xshift=1cm, align=left] {$\sim$: simulation invariant\\ $\ast$: 0 or several steps\\$\downarrow$: mixed semantics step\\\dashdownarrow: x86 or RTL step};
      \path [draw] (rtl1) edge[->, left] node {\footnotesize 4.\\\footnotesize silent~$\ast$} (rtl2);
      \path [draw] (rtl2) edge[->, left] node {\footnotesize 5.\\\footnotesize silent} (rtl3);
      \path [draw] (rtl3) edge[->, left] node {\footnotesize 4.\\\footnotesize silent~$\ast$} (rtl4);
      \path [draw] (x861) edge[->, right] node {\footnotesize 1.\\\footnotesize silent} (x862);
      \path [draw, shorten <=0.2cm, shorten >=0.2cm] ([xshift=-0.5cm]rtl1.east) edge[->, dashed, right] node {\footnotesize 3.\\\footnotesize$\ast$~silent} ([xshift=-0.5cm] rtl2.east);
      \path [draw, shorten <=0.2cm, shorten >=0.2cm] ([xshift=-0.5cm]rtl2.east) edge[->, dashed, right] node {\footnotesize 3.\\\footnotesize\coqin{Push}} ([xshift=-0.5cm] rtl3.east);
      \path [draw, shorten <=0.2cm, shorten >=0.2cm] ([xshift=-0.5cm]rtl3.east) edge[->, dashed, right] node {\footnotesize 3.\\\footnotesize$\ast$~silent} ([xshift=-0.5cm] rtl4.east);
      \path [draw, shorten <=0.2cm, shorten >=0.2cm] ([xshift=0.5cm]x861.west) edge[->, dashed, left] node {\footnotesize 2.\\\footnotesize\coqin{Push}} ([xshift=0.5cm] x862.west);
      \path [draw] (rtl1.east) edge[-, above] node {$\sim$} (x861.west);
      \path [draw] (rtl4.east) edge[-, below, sloped, bend right=12] node {$\sim$} (x862.south west);
    \end{tikzpicture}
    \caption{Preserving a primitive call in the mixed semantics.}
    \label{fig:backend_proof}
\end{figure}
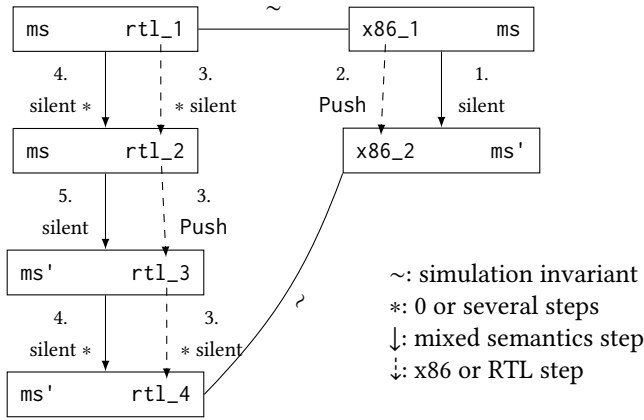

To prove correct the pass that uses the CompCert backend to transform RTL into x86, most of the effort lies in reusing CompCert simulations, relating x86 and RTL semantics, to construct a backward simulation on mixed semantics.
We first define a simulation invariant that relates two states of the mixed semantics $(rtls, ms)$ and $(x86s, ms)$ where $rtls$ and $x86s$ are semantic states of respectively RTL and x86, related by a CompCert simulation.
Since the programs have the same effects, the monadic state $ms$ must be identical.
Next, we prove a backward simulation: every semantic step of the mixed semantics after calling the CompCert backend is related to some steps of the mixed semantics before calling the backend.
Because the mixed semantics combine x86 and RTL semantics, we can reuse CompCert simulations again.

For instance, Figure~\ref{fig:backend_proof} illustrates one of the interesting cases of the proof: when the mixed step (1.) of the compiled program is calling the \coqin{Push} primitive (rule (push~x86) of Figure~\ref{fig:mixed_sem}).
We prove that this corresponds to a step (2.) with an observable behavior in the x86 semantics.
Then, using the CompCert backward simulation, we know that this step is matched by some steps in RTL semantics.
These RTL steps emit the same behavior and can be split into three parts (3.) around the non-silent step.
We  prove that silent RTL steps correspond to silent mixed steps (4.), and
that the RTL call at state \coqin{rtl_2} corresponds to a silent mixed step (5.) where the monadic effect of \coqin{Push} has been applied, turning \coqin{ms} into \coqin{ms'} just like it did on the x86 side. Finally, we have proved that the single silent step on the right can be matched with steps on the left that have the same monadic effect and thus preserve the invariant.
We then prove the entire backward simulation on mixed semantics. Outside of the compiled code, the invariant is simply reflexive.

\subsection{Putting Pieces Together: Main Correctness Theorem of \JITplain}
\label{subsec:jit_correct}
\outline{Theorem: backward simulation + behavior (see section2).
We reuse the nested simulation principle.
We need a backward for the entire optimization step: compose the simulations of the last two sections.
It's here that we prove that the JIT free monad corresponds to the mixed semantics interleaved with optimizations.}

\begin{figure}
  \centering
\begin{tikzpicture}[%
        every node/.style={rectangle,minimum size=0pt,minimum height=4pt, minimum width=30pt, inner sep=5pt, transform shape},
        shorten >=2pt,
        node distance=1cm, >=latex,
        align=center,
        scale = 0.95
      ]
      \node [] (input) [draw] {\IRplain semantics};
      \node [] (mixed) [draw, right=of input] {Mixed semantics};
      \node [] (jitref) [draw, right=of mixed] {JIT semantics\\\textit{Reference Spec}};
      \node [] (jitarray) [draw, right=of jitref] {JIT semantics\\\textit{Prim Spec}};
      \node [] (p) [draw, above=of mixed, xshift=-0.2cm] {$p$ \\\phantom{RTL}};
      \node [] (prtl) [draw, right=of p] {$p$ +\\ RTL};
      \node [] (px86) [draw, right=of prtl] {$p$ +\\x86};
      \node [] (opt) [left=of p] {Optimizer Correctness:};
      \node [] (jit) [below=of opt, yshift=0.5cm, xshift=-0.5cm] {JIT Correctness:};
      \path [draw] (mixed) edge[->, above, bend left] node {\footnotesize 1.} (input);
      \path [draw] (jitref) edge[->, above, bend left] node[midway](N) {\footnotesize 2.} (mixed);
      \path [draw] (jitref) edge[->, above, bend left] node {\footnotesize 3.} (jitarray);
      \path [draw] (jitarray) edge[->, above, bend left] node {\footnotesize 4.} (jitref);
      \path [draw] (p) edge[->, above, bend left] node {\footnotesize Section~\ref{subsec:callconv}} (prtl);
      \path [draw] (prtl) edge[->, above, bend left] node {\footnotesize Section~\ref{subsec:backend_correct}} (px86);
      \path [draw] (px86) edge[->, below, bend left] node[midway](M) {} (p);
      \path [draw] (M.center) edge[->, dashed] node {} ([xshift=0.3cm] N.south);
    \end{tikzpicture}
    \caption{Composing Simulations for the JIT Correctness Theorem.}
    \label{fig:jit_proof}
\end{figure}
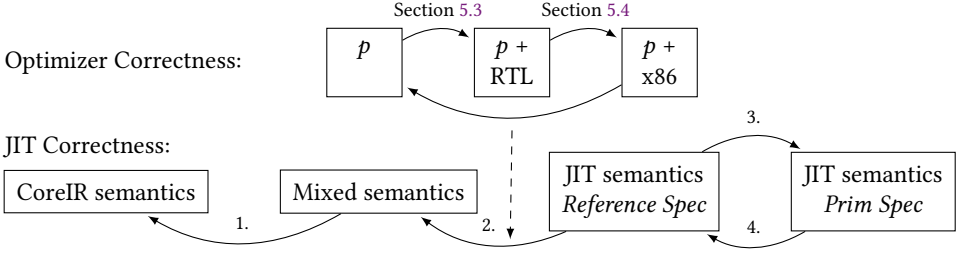

Finally, we prove the entire JIT correct by relating its semantics \coqin{jit_sem} to those of its original \IRplain program.
To do so, we write three more backward simulations.
Figure~\ref{fig:jit_proof} illustrates the different composed simulations in our proof, where right-facing arrow represent forward simulations, and left-facing arrow represent backward simulations.
First (1.), we prove that for any \IRplain program, its \IRplain semantics and its mixed semantics are simulated.
\IRplain semantics use neither free monads nor the global data-structures of \JITplain. On such a program with no native code yet, it suffices to prove that using the JIT stack and heap on a \IRplain-only program is equivalent.

Then (2.), we prove that for any program, its mixed semantics and its JIT semantics using the reference monadic specification are backward simulated.
There are two main challenges in this proof.
First, one needs to prove that inserting dynamic optimizations does not change the behavior.
This is done by reusing the nested simulation technique of~\cite{corejit}.
In a few words, inserting optimizations dynamically in the execution of a program can be proved with a backward simulation as long as this optimization step is proved correct with a backward simulation.
Our optimizer proof purposely fits this criteria, as we compose the two simulations of Sections~\ref{subsec:callconv} and~\ref{subsec:backend_correct}.
Second, we prove that the JIT semantics correspond to the mixed semantics interleaved with optimizations.
This entails relating states of the mixed semantics (Figure~\ref{fig:mixed_states}) to states of the JIT state-machine (Figure~\ref{fig:monitor}).
This also requires proving that the extended semantics of Section~\ref{subsec:nonatomic} for non-atomicity match the behavior of the mixed semantics when calling x86 code.
To assemble these two arguments in practice, we define other intermediate semantics and compose two backward simulations.

All of these simulations are written for the reference specification, where the JIT stack is structured.
We then use the refinement theorem of Section~\ref{subsec:refinement} to prove a forward simulation (3.) between the JIT semantics using the reference specification and the JIT semantics using \coqin{prim_spec}, a monadic specification much closer to the actual impure C primitive implementations.
This is a forward simulation that can be transformed into a backward simulation (4.) since the JIT behavior is deterministic.

Composing all these backward simulations, we get the \coqin{jit_correctness} theorem of Section~\ref{subsec:theorem}. Every JIT behavior using \coqin{prim_spec} of a program \coqin{p} refines a behavior of the \IRplain semantics of \coqin{p}.

\section{Assessment of \JITplain and Perspectives}\label{sec:implem}

This section first explains how to run \JITplain. Then, it discusses some of our design choices and some optimizations left as future work.
Last, the section quantifies our formal development and our proof effort, before detailing our trusted code base.

\subsection{Running the Executable \JITplain}
\outline{Compare the JIT to only using an interpreter.
No benchmarks.
Suggest optimizations: direct calls.}

We wrote a C implementation for all the JIT primitives, that the OCaml free interpreter of Section~\ref{subsec:impure_prims} can call.
Each of the stack and heap primitives uses a global array of 64-bit integers, and resembles its monadic specification (see Figure~\ref{fig:impl_and_spec}).
The primitives to install code are more involved and use system calls. The implementation of \coqin{Install_Code} allocates memory with \texttt{mmap}, writes into it after assembling the output of CompCert, then makes it executable with \texttt{mprotect}. We use the \texttt{elf} library\footnote{\url{https://man7.org/linux/man-pages/man5/elf.5.html}} to get the binary code of the assembled file.

After writing a parser for \IRplain and simple profiler heuristics (optimizing a function after a given number of calls), we can run \JITplain on actual programs.
This helps us validate experimentally that \JITplain is executable and that primitive implementations behave as expected.
Modern JIT compilers generate native code dynamically to execute faster than simply using an interpreter and this experiment shows that this is true of \JITplain as well.
On some programs, we observe up to 40 times speedups when dynamically compiling hot functions, compared to running our extracted interpreter alone. This is the case for instance for a naive program searching for prime numbers, when the function that uses a loop to check if a number is prime gets compiled.
Of course, these speedups are only as good as the profiler and real programs may require better heuristics.
When compiling a function that is barely called after, we can observe an execution time overhead due to calling the optimizer.
We have tested successfully \JITplain on all the features of \IRplain, including programs with speculation and deoptimization.
These examples programs are included in the artifact.

\subsection{Design Choices and Possible \JITplain Optimizations}
The design space of modern JIT compilers is particularly large, and ungoverned by any standard.
For instance, some JITs compile entire functions like we do (HotSpot), and others compile execution traces (TraceMonkey).
Some JITs use other IRs, like SSA or Sea-of-Nodes, that extend RTL-like languages with richer properties.
This is not a JIT-specific issue, and there exist previous works on using other IRs in a verified static compiler~\cite{compcert_ssa}.
The main drive behind the design choices of \JITplain is to define a simple yet representative architecture that captures JIT-specific behaviors (like interleaving native code and interpretation) that have yet to be formalized.

One may wonder if our synchronization interface, going back and from the monitor at each function call, can be a bottleneck for execution.
In~\cite{corejit}, CoreJIT used an unverified optimization when compiling a call to another already compiled function: the LLVM backend generated a direct call to that function.
This means that going back to the monitor is only needed when going to the interpreter is needed, and execution can stay at the native level as long as possible.
We believe that, as future work, a similar optimization asking CompCert to compile function calls for compiled functions, could be done and verified using \JITplain.
This is a benefit of having formalized and mechanized native code generation in a JIT: optimizations whose correctness depends on several components (the backend and the monitor) are now possible to prove.

Similarly, using external calls in the native code for each stack and heap interaction could be detrimental to execution times.
These external calls define a clear interface of the impure effects of the JIT.
One could now imagine possible optimizations, either inlining x86 implementations of the primitives as an additional step of the backend, or defining custom builtin functions at the RTL level that CompCert can also inline and compile.
Having defined semantics for native code execution in a JIT compiler, such optimizations could be proved correct in \JITplain.

\subsection{Proof Reuse}
\outline{CompCert simulation framework. CompCert backend proof. CompCert behavior proof: we do not have to write CoInductive proofs
Our proofs only target the JIT-specific issues.}

Our correctness proof successfully reuses a lot from CompCert.
The simulation framework is used for the final JIT theorem.
Liveness analysis used when generating RTL is done using the Kildall library of CompCert.
The behavior theorem that goes from a backward simulation to the final theorem of Section~\ref{subsec:theorem} also comes from CompCert and is the only proof using coinduction.

The entire Coq development represents around 15K lines we wrote, and includes more than 160K lines from CompCert.
This proof reuse allowed our proofs to entirely focus on JIT-specific issues.
We also wrote around 1K lines of OCaml for the free interpreter, profiling, parsing, pretty-printing and interfacing with C primitives. We include around 46K OCaml lines from CompCert.
Our C library is smaller than 500 lines.

\subsection{Trusted Code Base}
\label{subsec:tcb}
\outline{Coq extraction. Primitive Specification and Implementations. The loop and small-step semantics. The native call specification (non-atomicity).}

While most of the JIT is proved correct in Coq, there is still some code that has to be trusted in order to fully trust the OCaml executable JIT:
\begin{itemize}
\item The Coq to OCaml extraction of the free JIT.
\item The OCaml function that loops the JIT step should correspond to the JIT semantics. This is a just a few lines of code repeatedly calling the free interpreter until reaching a final state.
\item The free interpreter of Section~\ref{subsec:impure_prims}. This calls C functions from OCaml.\footnote{Currently, this uses \coqin{Obj.magic}. We could avoid that if Coq extraction could generate a GADT for the \coqin{primitive} type.}
\item The primitive implementations should correspond to their monadic specifications (Figure~\ref{fig:impl_and_spec}).
\item The call to native code should correspond to the three monadic specifications \coqin{start_}, \coqin{step_} and \coqin{end_} of Section~\ref{subsec:nonatomic}.
\end{itemize}

The task of verifying that the primitive implementations comply with their monadic specifications is out of scope of this paper.
In our work, we prove correct exactly the parts of the JIT that are extracted to OCaml, given a specification of the rest.
This orthogonal verification work could be done with the help of Hoare logic as well, for instance using VST~\cite{vst}.

First, note that when implementations deviate too much from their specification, one could use the refinement methodology to define another specification, closer to the C implementations.
However, there are a still few examples where the monadic specifications may not exactly match up with their implementations.
First, even in the \coqin{prim_spec} specification, the stack is assumed to be infinite.
In practice, we have implemented it as a finite array of 64-bits integers.
This is not a JIT-specific issue, and CompCert also assumes to be working with an infinite memory.
There exists a CompCert variant~\cite{stack_aware} that allows reasoning about bounded stack usage that we could investigate in a JIT setting as future work.

Second, in the formal model, the native code is represented by its x86 AST, just like in CompCert.
In practice however, we call an assembler to generate machine bytes for the native code. These machine bytes are installed in the memory, not the x86 AST.
This is not a JIT-specific issue, and we decided to go as far as CompCert goes in our formal model, 
at the cost of trusting the assembler.

Also, in its monadic specifications, the primitive that installs code never fails.
In practice however, our calls to \coqin{mmap} could fail if we ran out of memory to install the dynamically generated code.
We could solve this issue by allowing the primitive specification to non-deterministically fail and in such cases cancel the optimization step.
In our experiments so far, we have never encountered this issue.
Finally, we also need to trust that calling the native code is correctly specified with the three monads of Section~\ref{subsec:nonatomic}, \coqin{start_}, \coqin{step_} and \coqin{end_}.
The \coqin{step_} function simply reuses the CompCert x86 semantics, with an exception for primitives as seen on Figure~\ref{fig:mixed_sem}. In our experiments, we did not find any bugs with the execution of native code linked with primitives.


The CompCert theorem only holds for complete programs, where every piece of code has been compiled as a whole. This is not the case of the functions compiled by \JITplain.
We add three simple axioms in our Coq development to still reuse the CompCert theorem.
The first one strengthens an existing axiom of CompCert, specifying that calls to our JIT primitives have a precise annotated behavior in CompCert semantics, leaving unchanged the memory handled by CompCert.
We also assume that for each primitive and compiled function, there exists a place in CompCert memory where they have been allocated.
In practice, primitives have been compiled outside of the memory modeled by CompCert, as part of the JIT C library, but without this assumption the RTL programs we produce would have no semantics.
However, we then edit the generated code by CompCert to jump to the actual addresses of the primitives.
These two axioms are reminiscent of the way CompCert has dealt with some helper functions for integer arithmetics. 
These axioms are not incomplete proofs, but consequences of the need of CompCert to model a view of the complete memory.
Extending CompCert to support separate compilation, where several programs have different views of the memory, requires new proof techniques~\cite{compcomp,SongCKKKH20}.

Last, we include a simple axiom that could be proved by unfolding CompCert code transformations:
the CompCert backend does not generate any new built-ins call that is not already in the RTL programs we create. This was validated in our experiments.

\section{Related Work}\label{sec:sota}
\outline{Granularity (There is no standard currently. Functions is good to reuse CompCert, some JITs do it as well).}
\outline{Speculation (compatible with our previous work).}
\outline{Why this Intermediate Representation}

\subsection{Formally Verified JITs}
Other works have tackled the issue of formalizing some parts of modern JITs.
Focusing on the range analysis Javascript JITs perform, a DSL that uses an SMT solver to write and prove the correctness of range analysis in JITs has been developed in~\cite{range_jit}.
Focusing on speculations, semantics preservation proofs for speculative optimizations were first studied in the Sourir intermediate representation~\cite{sourir}.
Sourir introduced formal semantics for speculative instructions like our \textbf{Assume}, and arguments for the correctness of inserting and manipulating them.
CoreJIT~\cite{corejit} mechanized these formal semantics and verified in Coq dynamic code transformations inserting and manipulating speculative instructions.
CoreJIT itself was a prototype of a JIT compiler using \IRplain, with only interpretation and a speculative dynamic optimizer, from \IRplain to \IRplain.
This prototype could also be extracted to OCaml and was completed with an unverified and unspecified LLVM backend compiler.
\JITplain shares several similarities with CoreJIT.
Both purposely share a similar architecture, and both require their dynamic optimizations to be proved correct with a backward simulation.
As such simulations compose, one could easily imagine implementing the speculative optimizations of CoreJIT in \JITplain before the backend, at the only cost of adapting the proofs to the monadic semantics.
However, CoreJIT did not formalize the generation of native code, its LLVM backend was not part of the formal model, and
there was no proof about the interaction between IR interpretation and native execution.
Modern JITs rely greatly on that interplay which remained, until \JITplain, an open verification problem.
The CoreJIT model also restricted itself to a pure subset of JIT components, with no shared data-structure, no code installation, no call to native code and with no insight for the verification of a realistic effectful JIT.
We believe that \JITplain has successfully tackled these remaining challenges.
In the end, \JITplain and CoreJIT target different and complementary essential verification issues of modern JIT compilers, using compatible designs.

\cite{myreen_jit} is another work of JIT formal verification for a stack-based bytecode, which in particular targets the challenge of dynamically generating x86 code.
Proofs are mechanized with HOL4, and the result is an executable JIT compiler which dynamically generates native code for each called function.
To that end, this work defines semantics for self-modifying x86 code.
However, this JIT workflow is not characteristic of modern JIT compilers that interleave multiple tiers of execution (interpretation and native code execution).
There are no speculations and without an interpreter, no on-stack-replacement.
We believe that \JITplain proposes a design more typical of modern JIT compilers, enabling reasoning about their precise interoperability.

\subsection{Verifying Effectful Programs in Coq}
Our work is not the first Coq mechanization about effectful programs.
There are various ways to go around the limitations of Gallina as a programming language.
For instance, a monadic approach is followed with Isabelle/HOL for the verified
seL4 microkernel~\cite{CockKS08}, in order to refine monadic
functional specifications into a C implementation. Moreover,
\cite{Pit-ClaudelWDGC20} avoid the extraction from Coq to OCaml to
produce effectful verified programs. They directly transform a functional
specification into a fully linked assembly program represented as a Coq term.
Modifying Coq extraction so that it can produce effectful OCaml programs has also been investigated for programs using mutable arrays~\cite{extraction_arrays}.
Programs are written using a state monad encoding, and the improved extraction produces efficient OCaml code using mutable data-structures.
In contrast, our formalism allows us to use extraction to OCaml without modifications, but only translating the pure parts of the program.

Other contributions have explored using variations of free monads before us, but our version was designed to be lightweight, specialized to our JIT and compatible with CompCert.
FreeSpec~\cite{freespec} uses a free monadic definition similar to ours to encode programs with effects.
Their definition is more general in the sense that one can compose several interfaces of primitives, while our single interface of primitives is specialized to those used by the JIT.
Our monadic definitions mainly differ in the way primitives are specified.
We use state monads allows to simply define small-step operational semantics {\it à la} CompCert.

Interaction Trees~\cite{itrees} are a coinductive variant of free monads; they are used in~\cite{itrees_llvm} to define the semantics of a subset of LLVM.
Like in our approach, computations using interaction trees can be extracted to OCaml and executed using effectful event handlers, much like our \coqin{free_interpreter}, where events include the calls to effectful primitives.
Their use of a coinductive structure allows interaction trees to represent diverging computations with some added monadic constructors, at the only cost of developing a library for coinductive reasoning.
Coinductive reasoning can be difficult to work with in Coq.
We entirely avoid this issue by breaking down what a JIT does in small, atomic computations (the transitions of Figure~\ref{fig:monitor}). Even possibly diverging transitions can be broken down themselves in small steps (\figref{unf_sem}).
Finally, our lightweight monadic library does not have the slightest coinductive proof but we rather reuse one of CompCert, going from a simulation to the \coqin{jit_correctness} theorem of Section~\ref{subsec:theorem}.
Interaction Trees are an impressive framework for the verification of effectful diverging programs, but in the specific case of a formally verified JIT, there is much to gain by staying close to CompCert and its proof techniques to reuse its correctness proof.

Our refinement methodology is also reminiscent of other works using refinement to prove the correctness of a concrete implementation using a more convenient abstraction.
For instance, Isabelle includes a refinement framework which has been used to verify network flow algorithms~\cite{refinement_flow_isabelle}.
In contrast, our refinement methodology is designed specifically for the novel interaction between the free monad formalism and the CompCert simulation framework.
Our refinement definition purposely resembles a CompCert forward simulation.

\section{Conclusion}\label{sec:concl}
\outline{Verified and executable.
Clear specification of each component.  Some parts cannot be written
in Coq: we wrote exactly what we wanted to extract and proved that
correct.}

JIT compilers are complex pieces of software relying on cutting edge techniques. Not only do they generate native code, like static compilers, but they also include an entire execution environment with various components.
Modern JITs have been scarcely formalized, and are in dire need of demystification if one ever wants strong guarantees on the execution of a JIT.
The need for such guarantees is made more crucial by the adoption of JITs in many critical environments,
 including to run possibly adversarial programs as in most of modern Web browsers.

While there is no general agreement on the components of a JIT and their interplay, we believe that \JITplain is a reasonable proposal that works well in the context of compiler verification.
This mechanized JIT in Coq captures the essence of a JIT with native code generation, where each component is specified and proved correct.
We demonstrate that, just like JIT compilers reuse static compilation techniques, formally verified JIT compilers can reuse formally verified static compilers like CompCert.
Writing a JIT with native code generation entirely in Coq is impossible due to its intrinsically effectful nature.
However, we were able to design a JIT where effects are delimited to a very restricted set of primitives that are specified in Coq.
This methodology also allows us to extract \JITplain to OCaml and complete it with effectful implementations of the primitives. In the end, we have a JIT that is both executable and formally verified in Coq.

With these essential JIT-specific issues solved, one can now imagine many ways to go forward in verifying more realistic JIT compilers.
An interesting and short-term one would be to extend \JITplain with the speculation insertion of CoreJIT~\cite{corejit}, now that speculative instruction compilation is provably feasible.
Another direction for future work is to extend our JIT to a more realistic input language, such as WebAssembly~\cite{wasm_mechanized} which already has a semantics mechanized in Coq.


\bibliographystyle{ACM-Reference-Format}
\bibliography{bibliography}


\begin{thebibliography}{34}


\ifx \showCODEN    \undefined \def \showCODEN     #1{\unskip}     \fi
\ifx \showDOI      \undefined \def \showDOI       #1{#1}\fi
\ifx \showISBNx    \undefined \def \showISBNx     #1{\unskip}     \fi
\ifx \showISBNxiii \undefined \def \showISBNxiii  #1{\unskip}     \fi
\ifx \showISSN     \undefined \def \showISSN      #1{\unskip}     \fi
\ifx \showLCCN     \undefined \def \showLCCN      #1{\unskip}     \fi
\ifx \shownote     \undefined \def \shownote      #1{#1}          \fi
\ifx \showarticletitle \undefined \def \showarticletitle #1{#1}   \fi
\ifx \showURL      \undefined \def \showURL       {\relax}        \fi
\providecommand\bibfield[2]{#2}
\providecommand\bibinfo[2]{#2}
\providecommand\natexlab[1]{#1}
\providecommand\showeprint[2][]{arXiv:#2}

\bibitem[\protect\citeauthoryear{Appel}{Appel}{2015}]%
        {vst}
\bibfield{author}{\bibinfo{person}{Andrew~W. Appel}.}
  \bibinfo{year}{2015}\natexlab{}.
\newblock \showarticletitle{Verification of a Cryptographic Primitive:
  {SHA-256}}.
\newblock \bibinfo{journal}{\emph{{ACM} Trans. Program. Lang. Syst.}}
  \bibinfo{volume}{37}, \bibinfo{number}{2} (\bibinfo{year}{2015}),
  \bibinfo{pages}{7:1--7:31}.
\newblock
\urldef\tempurl%
\url{https://doi.org/10.1145/2701415}
\showDOI{\tempurl}


\bibitem[\protect\citeauthoryear{Barri{\`{e}}re, Blazy, Fl{\"{u}}ckiger,
  Pichardie, and Vitek}{Barri{\`{e}}re et~al\mbox{.}}{2021}]%
        {corejit}
\bibfield{author}{\bibinfo{person}{Aur{\`{e}}le Barri{\`{e}}re},
  \bibinfo{person}{Sandrine Blazy}, \bibinfo{person}{Olivier Fl{\"{u}}ckiger},
  \bibinfo{person}{David Pichardie}, {and} \bibinfo{person}{Jan Vitek}.}
  \bibinfo{year}{2021}\natexlab{}.
\newblock \showarticletitle{Formally verified speculation and deoptimization in
  a {JIT} compiler}.
\newblock \bibinfo{journal}{\emph{Proc. {ACM} Program. Lang.}}
  \bibinfo{number}{{POPL}} (\bibinfo{year}{2021}).
\newblock
\urldef\tempurl%
\url{https://doi.org/10.1145/3434327}
\showURL{%
\tempurl}


\bibitem[\protect\citeauthoryear{Barthe, Demange, and Pichardie}{Barthe
  et~al\mbox{.}}{2014}]%
        {compcert_ssa}
\bibfield{author}{\bibinfo{person}{Gilles Barthe}, \bibinfo{person}{Delphine
  Demange}, {and} \bibinfo{person}{David Pichardie}.}
  \bibinfo{year}{2014}\natexlab{}.
\newblock \showarticletitle{Formal Verification of an SSA-Based Middle-End for
  CompCert}.
\newblock \bibinfo{journal}{\emph{{ACM} Trans. Program. Lang. Syst.}}
  \bibinfo{volume}{36}, \bibinfo{number}{1} (\bibinfo{year}{2014}),
  \bibinfo{pages}{4:1--4:35}.
\newblock
\urldef\tempurl%
\url{https://doi.org/10.1145/2579080}
\showDOI{\tempurl}


\bibitem[\protect\citeauthoryear{Brown, Renner, N{\"{o}}tzli, Lerner, Shacham,
  and Stefan}{Brown et~al\mbox{.}}{2020}]%
        {range_jit}
\bibfield{author}{\bibinfo{person}{Fraser Brown}, \bibinfo{person}{John
  Renner}, \bibinfo{person}{Andres N{\"{o}}tzli}, \bibinfo{person}{Sorin
  Lerner}, \bibinfo{person}{Hovav Shacham}, {and} \bibinfo{person}{Deian
  Stefan}.} \bibinfo{year}{2020}\natexlab{}.
\newblock \showarticletitle{Towards a verified range analysis for JavaScript
  JITs}. In \bibinfo{booktitle}{\emph{Proceedings of the 41st {ACM} {SIGPLAN}
  International Conference on Programming Language Design and Implementation,
  {PLDI} 2020}}. \bibinfo{publisher}{{ACM}}, \bibinfo{pages}{135--150}.
\newblock
\urldef\tempurl%
\url{https://doi.org/10.1145/3385412.3385968}
\showDOI{\tempurl}


\bibitem[\protect\citeauthoryear{Cock, Klein, and Sewell}{Cock
  et~al\mbox{.}}{2008}]%
        {CockKS08}
\bibfield{author}{\bibinfo{person}{David Cock}, \bibinfo{person}{Gerwin Klein},
  {and} \bibinfo{person}{Thomas Sewell}.} \bibinfo{year}{2008}\natexlab{}.
\newblock \showarticletitle{Secure Microkernels, State Monads and Scalable
  Refinement}. In \bibinfo{booktitle}{\emph{Proc. of {TPHOLs} 2008}},
  Vol.~\bibinfo{volume}{5170}. \bibinfo{publisher}{Springer},
  \bibinfo{pages}{167--182}.
\newblock
\urldef\tempurl%
\url{https://doi.org/10.1007/978-3-540-71067-7\_16}
\showDOI{\tempurl}


\bibitem[\protect\citeauthoryear{Fl{\"{u}}ckiger, Scherer, Yee, Goel, Ahmed,
  and Vitek}{Fl{\"{u}}ckiger et~al\mbox{.}}{2018}]%
        {sourir}
\bibfield{author}{\bibinfo{person}{Olivier Fl{\"{u}}ckiger},
  \bibinfo{person}{Gabriel Scherer}, \bibinfo{person}{Ming{-}Ho Yee},
  \bibinfo{person}{Aviral Goel}, \bibinfo{person}{Amal Ahmed}, {and}
  \bibinfo{person}{Jan Vitek}.} \bibinfo{year}{2018}\natexlab{}.
\newblock \showarticletitle{Correctness of speculative optimizations with
  dynamic deoptimization}.
\newblock  \bibinfo{number}{{POPL}} (\bibinfo{year}{2018}).
\newblock
\urldef\tempurl%
\url{https://doi.org/10.1145/3158137}
\showDOI{\tempurl}


\bibitem[\protect\citeauthoryear{HotSpot}{HotSpot}{2022}]%
        {hotspot}
HotSpot \bibinfo{year}{2022}\natexlab{}.
\newblock \bibinfo{booktitle}{\emph{Java {HotSpot} Performance Engine}}.
\newblock HotSpot.
\newblock
\urldef\tempurl%
\url{https://openjdk.org/groups/hotspot/}
\showURL{%
\tempurl}


\bibitem[\protect\citeauthoryear{Inria}{Inria}{2022}]%
        {Coq}
Inria \bibinfo{year}{2022}\natexlab{}.
\newblock \bibinfo{booktitle}{\emph{The {Coq} proof assistant reference
  manual}}.
\newblock Inria.
\newblock
\urldef\tempurl%
\url{http://coq.inria.fr}
\showURL{%
\tempurl}
\newblock
\shownote{Version 8.12.1.}


\bibitem[\protect\citeauthoryear{Kang, Kim, Song, Lee, Park, Shin, Kim, Cho,
  Choi, Hur, and Yi}{Kang et~al\mbox{.}}{2018}]%
        {KangKSLPSKCCHY18}
\bibfield{author}{\bibinfo{person}{Jeehoon Kang}, \bibinfo{person}{Yoonseung
  Kim}, \bibinfo{person}{Youngju Song}, \bibinfo{person}{Juneyoung Lee},
  \bibinfo{person}{Sanghoon Park}, \bibinfo{person}{Mark~Dongyeon Shin},
  \bibinfo{person}{Yonghyun Kim}, \bibinfo{person}{Sungkeun Cho},
  \bibinfo{person}{Joonwon Choi}, \bibinfo{person}{Chung{-}Kil Hur}, {and}
  \bibinfo{person}{Kwangkeun Yi}.} \bibinfo{year}{2018}\natexlab{}.
\newblock \showarticletitle{Crellvm: verified credible compilation for {LLVM}}.
  In \bibinfo{booktitle}{\emph{Proceedings of the 39th {ACM} {SIGPLAN}
  Conference on Programming Language Design and Implementation, {PLDI} 2018}}.
  \bibinfo{publisher}{{ACM}}, \bibinfo{pages}{631--645}.
\newblock
\urldef\tempurl%
\url{https://doi.org/10.1145/3192366.3192377}
\showDOI{\tempurl}


\bibitem[\protect\citeauthoryear{K{\"a}stner, Barrho, W{\"u}nsche, Schlickling,
  Schommer, Schmidt, Ferdinand, Leroy, and Blazy}{K{\"a}stner
  et~al\mbox{.}}{2018}]%
        {kastner:hal-01643290}
\bibfield{author}{\bibinfo{person}{Daniel K{\"a}stner},
  \bibinfo{person}{J{\"o}rg Barrho}, \bibinfo{person}{Ulrich W{\"u}nsche},
  \bibinfo{person}{Marc Schlickling}, \bibinfo{person}{Bernhard Schommer},
  \bibinfo{person}{Michael Schmidt}, \bibinfo{person}{Christian Ferdinand},
  \bibinfo{person}{Xavier Leroy}, {and} \bibinfo{person}{Sandrine Blazy}.}
  \bibinfo{year}{2018}\natexlab{}.
\newblock \showarticletitle{{CompCert: Practical Experience on Integrating and
  Qualifying a Formally Verified Optimizing Compiler}}. In
  \bibinfo{booktitle}{\emph{{ERTS2 2018 - 9th European Congress Embedded
  Real-Time Software and Systems}}}. {3AF, SEE, SIE}, \bibinfo{pages}{1--9}.
\newblock
\urldef\tempurl%
\url{https://hal.inria.fr/hal-01643290}
\showURL{%
\tempurl}


\bibitem[\protect\citeauthoryear{Kumar, Myreen, Norrish, and Owens}{Kumar
  et~al\mbox{.}}{2014}]%
        {cakeml}
\bibfield{author}{\bibinfo{person}{Ramana Kumar}, \bibinfo{person}{Magnus~O.
  Myreen}, \bibinfo{person}{Michael Norrish}, {and} \bibinfo{person}{Scott
  Owens}.} \bibinfo{year}{2014}\natexlab{}.
\newblock \showarticletitle{CakeML: a verified implementation of {ML}}. In
  \bibinfo{booktitle}{\emph{Proceedings of {POPL}}}.
\newblock
\urldef\tempurl%
\url{https://doi.org/10.1145/2535838.2535841}
\showDOI{\tempurl}


\bibitem[\protect\citeauthoryear{Lammich and Sefidgar}{Lammich and
  Sefidgar}{2019}]%
        {refinement_flow_isabelle}
\bibfield{author}{\bibinfo{person}{Peter Lammich} {and}
  \bibinfo{person}{S.~Reza Sefidgar}.} \bibinfo{year}{2019}\natexlab{}.
\newblock \showarticletitle{Formalizing Network Flow Algorithms: {A} Refinement
  Approach in Isabelle/HOL}.
\newblock \bibinfo{journal}{\emph{J. Autom. Reason.}} \bibinfo{volume}{62},
  \bibinfo{number}{2} (\bibinfo{year}{2019}), \bibinfo{pages}{261--280}.
\newblock
\urldef\tempurl%
\url{https://doi.org/10.1007/s10817-017-9442-4}
\showDOI{\tempurl}


\bibitem[\protect\citeauthoryear{Leroy}{Leroy}{2006}]%
        {compcert}
\bibfield{author}{\bibinfo{person}{Xavier Leroy}.}
  \bibinfo{year}{2006}\natexlab{}.
\newblock \showarticletitle{Formal certification of a compiler back-end or:
  programming a compiler with a proof assistant}. In
  \bibinfo{booktitle}{\emph{Proceedings of {POPL}}}.
\newblock
\urldef\tempurl%
\url{https://doi.org/10.1145/1111037.1111042}
\showDOI{\tempurl}


\bibitem[\protect\citeauthoryear{Leroy}{Leroy}{2009a}]%
        {CACM:compcert}
\bibfield{author}{\bibinfo{person}{Xavier Leroy}.}
  \bibinfo{year}{2009}\natexlab{a}.
\newblock \showarticletitle{Formal verification of a realistic compiler}.
\newblock \bibinfo{journal}{\emph{Commun. ACM}} (\bibinfo{year}{2009}).
\newblock
\urldef\tempurl%
\url{https://doi.org/10.1145/1538788.1538814}
\showDOI{\tempurl}


\bibitem[\protect\citeauthoryear{Leroy}{Leroy}{2009b}]%
        {Leroy-backend}
\bibfield{author}{\bibinfo{person}{Xavier Leroy}.}
  \bibinfo{year}{2009}\natexlab{b}.
\newblock \showarticletitle{A formally verified compiler back-end}.
\newblock \bibinfo{journal}{\emph{Journal of Automated Reasoning}}
  \bibinfo{volume}{43}, \bibinfo{number}{4} (\bibinfo{year}{2009}),
  \bibinfo{pages}{363--446}.
\newblock
\urldef\tempurl%
\url{https://doi.org/10.1007/s10817-009-9155-4}
\showDOI{\tempurl}


\bibitem[\protect\citeauthoryear{Leroy, Blazy, K{\"a}stner, Schommer, Pister,
  and Ferdinand}{Leroy et~al\mbox{.}}{2016}]%
        {leroy:hal-01238879}
\bibfield{author}{\bibinfo{person}{Xavier Leroy}, \bibinfo{person}{Sandrine
  Blazy}, \bibinfo{person}{Daniel K{\"a}stner}, \bibinfo{person}{Bernhard
  Schommer}, \bibinfo{person}{Markus Pister}, {and} \bibinfo{person}{Christian
  Ferdinand}.} \bibinfo{year}{2016}\natexlab{}.
\newblock \showarticletitle{{CompCert - A Formally Verified Optimizing
  Compiler}}. In \bibinfo{booktitle}{\emph{{ERTS 2016: Embedded Real Time
  Software and Systems, 8th European Congress}}}. {SEE}.
\newblock
\urldef\tempurl%
\url{https://hal.inria.fr/hal-01238879}
\showURL{%
\tempurl}


\bibitem[\protect\citeauthoryear{Letan and R{\'{e}}gis{-}Gianas}{Letan and
  R{\'{e}}gis{-}Gianas}{2020}]%
        {freespec}
\bibfield{author}{\bibinfo{person}{Thomas Letan} {and} \bibinfo{person}{Yann
  R{\'{e}}gis{-}Gianas}.} \bibinfo{year}{2020}\natexlab{}.
\newblock \showarticletitle{FreeSpec: specifying, verifying, and executing
  impure computations in Coq}. In \bibinfo{booktitle}{\emph{Proceedings of the
  9th {ACM} {SIGPLAN} International Conference on Certified Programs and
  Proofs, {CPP}}}.
\newblock
\urldef\tempurl%
\url{https://doi.org/10.1145/3372885.3373812}
\showURL{%
\tempurl}


\bibitem[\protect\citeauthoryear{L{\"{o}}{\"{o}}w, Kumar, Tan, Myreen, Norrish,
  Abrahamsson, and Fox}{L{\"{o}}{\"{o}}w et~al\mbox{.}}{2019}]%
        {LoowKTMNAF19}
\bibfield{author}{\bibinfo{person}{Andreas L{\"{o}}{\"{o}}w},
  \bibinfo{person}{Ramana Kumar}, \bibinfo{person}{Yong~Kiam Tan},
  \bibinfo{person}{Magnus~O. Myreen}, \bibinfo{person}{Michael Norrish},
  \bibinfo{person}{Oskar Abrahamsson}, {and} \bibinfo{person}{Anthony C.~J.
  Fox}.} \bibinfo{year}{2019}\natexlab{}.
\newblock \showarticletitle{Verified compilation on a verified processor}. In
  \bibinfo{booktitle}{\emph{Proceedings of the 40th {ACM} {SIGPLAN} Conference
  on Programming Language Design and Implementation, {PLDI} 2019}}.
  \bibinfo{publisher}{{ACM}}, \bibinfo{pages}{1041--1053}.
\newblock
\urldef\tempurl%
\url{https://doi.org/10.1145/3314221.3314622}
\showDOI{\tempurl}


\bibitem[\protect\citeauthoryear{Myreen}{Myreen}{2010}]%
        {myreen_jit}
\bibfield{author}{\bibinfo{person}{Magnus~O. Myreen}.}
  \bibinfo{year}{2010}\natexlab{}.
\newblock \showarticletitle{Verified just-in-time compiler on x86}. In
  \bibinfo{booktitle}{\emph{Proceedings of the 37th {ACM} {SIGPLAN-SIGACT}
  Symposium on Principles of Programming Languages, {POPL} 2010}}.
  \bibinfo{publisher}{{ACM}}, \bibinfo{pages}{107--118}.
\newblock
\urldef\tempurl%
\url{https://doi.org/10.1145/1706299.1706313}
\showDOI{\tempurl}


\bibitem[\protect\citeauthoryear{Owens, Norrish, Kumar, Myreen, and Tan}{Owens
  et~al\mbox{.}}{2017}]%
        {OwensNKMT17}
\bibfield{author}{\bibinfo{person}{Scott Owens}, \bibinfo{person}{Michael
  Norrish}, \bibinfo{person}{Ramana Kumar}, \bibinfo{person}{Magnus~O. Myreen},
  {and} \bibinfo{person}{Yong~Kiam Tan}.} \bibinfo{year}{2017}\natexlab{}.
\newblock \showarticletitle{Verifying efficient function calls in CakeML}.
\newblock \bibinfo{journal}{\emph{{PACMPL}}} \bibinfo{volume}{1},
  \bibinfo{number}{{ICFP}} (\bibinfo{year}{2017}),
  \bibinfo{pages}{18:1--18:27}.
\newblock
\urldef\tempurl%
\url{https://doi.org/10.1145/3110262}
\showDOI{\tempurl}


\bibitem[\protect\citeauthoryear{Pit{-}Claudel, Wang, Delaware, Gross, and
  Chlipala}{Pit{-}Claudel et~al\mbox{.}}{2020}]%
        {Pit-ClaudelWDGC20}
\bibfield{author}{\bibinfo{person}{Cl{\'{e}}ment Pit{-}Claudel},
  \bibinfo{person}{Peng Wang}, \bibinfo{person}{Benjamin Delaware},
  \bibinfo{person}{Jason Gross}, {and} \bibinfo{person}{Adam Chlipala}.}
  \bibinfo{year}{2020}\natexlab{}.
\newblock \showarticletitle{Extensible Extraction of Efficient Imperative
  Programs with Foreign Functions, Manually Managed Memory, and Proofs}. In
  \bibinfo{booktitle}{\emph{Proc. of {IJCAR} 2020}},
  Vol.~\bibinfo{volume}{12167}. \bibinfo{publisher}{Springer},
  \bibinfo{pages}{119--137}.
\newblock
\urldef\tempurl%
\url{https://doi.org/10.1007/978-3-030-51054-1\_7}
\showDOI{\tempurl}


\bibitem[\protect\citeauthoryear{PyPy}{PyPy}{2022}]%
        {pypy}
PyPy \bibinfo{year}{2022}\natexlab{}.
\newblock \bibinfo{booktitle}{\emph{{PyPy} Python Implementation}}.
\newblock PyPy.
\newblock
\urldef\tempurl%
\url{https://www.pypy.org/}
\showURL{%
\tempurl}


\bibitem[\protect\citeauthoryear{Sakaguchi}{Sakaguchi}{2018}]%
        {extraction_arrays}
\bibfield{author}{\bibinfo{person}{Kazuhiko Sakaguchi}.}
  \bibinfo{year}{2018}\natexlab{}.
\newblock \showarticletitle{Program Extraction for Mutable Arrays}. In
  \bibinfo{booktitle}{\emph{Functional and Logic Programming - 14th
  International Symposium, {FLOPS} 2018}}, Vol.~\bibinfo{volume}{10818}.
  \bibinfo{publisher}{Springer}, \bibinfo{pages}{51--67}.
\newblock
\urldef\tempurl%
\url{https://doi.org/10.1007/978-3-319-90686-7\_4}
\showDOI{\tempurl}


\bibitem[\protect\citeauthoryear{Song, Cho, Kim, Kim, Kang, and Hur}{Song
  et~al\mbox{.}}{2020}]%
        {SongCKKKH20}
\bibfield{author}{\bibinfo{person}{Youngju Song}, \bibinfo{person}{Minki Cho},
  \bibinfo{person}{Dongjoo Kim}, \bibinfo{person}{Yonghyun Kim},
  \bibinfo{person}{Jeehoon Kang}, {and} \bibinfo{person}{Chung{-}Kil Hur}.}
  \bibinfo{year}{2020}\natexlab{}.
\newblock \showarticletitle{CompCertM: CompCert with C-assembly linking and
  lightweight modular verification}.
\newblock \bibinfo{journal}{\emph{Proc. {ACM} Program. Lang.}}
  \bibinfo{volume}{4}, \bibinfo{number}{{POPL}} (\bibinfo{year}{2020}),
  \bibinfo{pages}{23:1--23:31}.
\newblock
\urldef\tempurl%
\url{https://doi.org/10.1145/3371091}
\showDOI{\tempurl}


\bibitem[\protect\citeauthoryear{Stewart, Beringer, Cuellar, and Appel}{Stewart
  et~al\mbox{.}}{2015}]%
        {compcomp}
\bibfield{author}{\bibinfo{person}{Gordon Stewart}, \bibinfo{person}{Lennart
  Beringer}, \bibinfo{person}{Santiago Cuellar}, {and}
  \bibinfo{person}{Andrew~W. Appel}.} \bibinfo{year}{2015}\natexlab{}.
\newblock \showarticletitle{Compositional CompCert}. In
  \bibinfo{booktitle}{\emph{Proc. {ACM} Program. Lang. {POPL} 2015}}.
  \bibinfo{publisher}{{ACM}}, \bibinfo{pages}{275--287}.
\newblock
\urldef\tempurl%
\url{https://doi.org/10.1145/2676726.2676985}
\showDOI{\tempurl}


\bibitem[\protect\citeauthoryear{Swierstra}{Swierstra}{2008}]%
        {datatypesalacarte}
\bibfield{author}{\bibinfo{person}{Wouter Swierstra}.}
  \bibinfo{year}{2008}\natexlab{}.
\newblock \showarticletitle{Data types {\`{a}} la carte}.
\newblock \bibinfo{journal}{\emph{J. Funct. Program.}} (\bibinfo{year}{2008}).
\newblock
\urldef\tempurl%
\url{https://doi.org/10.1017/S0956796808006758}
\showURL{%
\tempurl}


\bibitem[\protect\citeauthoryear{Tan, Myreen, Kumar, Fox, Owens, and
  Norrish}{Tan et~al\mbox{.}}{2016}]%
        {TanMKFON16}
\bibfield{author}{\bibinfo{person}{Yong~Kiam Tan}, \bibinfo{person}{Magnus~O.
  Myreen}, \bibinfo{person}{Ramana Kumar}, \bibinfo{person}{Anthony C.~J. Fox},
  \bibinfo{person}{Scott Owens}, {and} \bibinfo{person}{Michael Norrish}.}
  \bibinfo{year}{2016}\natexlab{}.
\newblock \showarticletitle{A new verified compiler backend for CakeML}. In
  \bibinfo{booktitle}{\emph{Proceedings of the 21st {ACM} {SIGPLAN}
  International Conference on Functional Programming, {ICFP} 2016}}.
  \bibinfo{publisher}{{ACM}}, \bibinfo{pages}{60--73}.
\newblock
\urldef\tempurl%
\url{https://doi.org/10.1145/2951913.2951924}
\showDOI{\tempurl}


\bibitem[\protect\citeauthoryear{V8}{V8}{2022}]%
        {v8}
V8 \bibinfo{year}{2022}\natexlab{}.
\newblock \bibinfo{booktitle}{\emph{{V8} Javascript Engine}}.
\newblock V8.
\newblock
\urldef\tempurl%
\url{https://v8.dev/}
\showURL{%
\tempurl}


\bibitem[\protect\citeauthoryear{Wang, Wilke, and Shao}{Wang
  et~al\mbox{.}}{2019}]%
        {stack_aware}
\bibfield{author}{\bibinfo{person}{Yuting Wang}, \bibinfo{person}{Pierre
  Wilke}, {and} \bibinfo{person}{Zhong Shao}.} \bibinfo{year}{2019}\natexlab{}.
\newblock \showarticletitle{An abstract stack based approach to verified
  compositional compilation to machine code}.
\newblock \bibinfo{journal}{\emph{Proc. {ACM} Program. Lang.}}
  \bibinfo{volume}{3}, \bibinfo{number}{{POPL}} (\bibinfo{year}{2019}),
  \bibinfo{pages}{62:1--62:30}.
\newblock
\urldef\tempurl%
\url{https://doi.org/10.1145/3290375}
\showDOI{\tempurl}


\bibitem[\protect\citeauthoryear{Watt, Rao, Pichon{-}Pharabod, Bodin, and
  Gardner}{Watt et~al\mbox{.}}{2021}]%
        {wasm_mechanized}
\bibfield{author}{\bibinfo{person}{Conrad Watt}, \bibinfo{person}{Xiaojia Rao},
  \bibinfo{person}{Jean Pichon{-}Pharabod}, \bibinfo{person}{Martin Bodin},
  {and} \bibinfo{person}{Philippa Gardner}.} \bibinfo{year}{2021}\natexlab{}.
\newblock \showarticletitle{Two Mechanisations of WebAssembly 1.0}. In
  \bibinfo{booktitle}{\emph{Formal Methods - 24th International Symposium, {FM}
  2021}}, Vol.~\bibinfo{volume}{13047}. \bibinfo{publisher}{Springer},
  \bibinfo{pages}{61--79}.
\newblock
\urldef\tempurl%
\url{https://doi.org/10.1007/978-3-030-90870-6\_4}
\showDOI{\tempurl}


\bibitem[\protect\citeauthoryear{Xia, Zakowski, He, Hur, Malecha, Pierce, and
  Zdancewic}{Xia et~al\mbox{.}}{2020}]%
        {itrees}
\bibfield{author}{\bibinfo{person}{Li{-}yao Xia}, \bibinfo{person}{Yannick
  Zakowski}, \bibinfo{person}{Paul He}, \bibinfo{person}{Chung{-}Kil Hur},
  \bibinfo{person}{Gregory Malecha}, \bibinfo{person}{Benjamin~C. Pierce},
  {and} \bibinfo{person}{Steve Zdancewic}.} \bibinfo{year}{2020}\natexlab{}.
\newblock \showarticletitle{Interaction trees: representing recursive and
  impure programs in Coq}.
\newblock \bibinfo{journal}{\emph{Proc. {ACM} Program. Lang.}}
  \bibinfo{number}{{POPL}} (\bibinfo{year}{2020}).
\newblock
\urldef\tempurl%
\url{https://doi.org/10.1145/3371119}
\showURL{%
\tempurl}


\bibitem[\protect\citeauthoryear{Zakowski, Beck, Yoon, Zaichuk, Zaliva, and
  Zdancewic}{Zakowski et~al\mbox{.}}{2021}]%
        {itrees_llvm}
\bibfield{author}{\bibinfo{person}{Yannick Zakowski}, \bibinfo{person}{Calvin
  Beck}, \bibinfo{person}{Irene Yoon}, \bibinfo{person}{Ilia Zaichuk},
  \bibinfo{person}{Vadim Zaliva}, {and} \bibinfo{person}{Steve Zdancewic}.}
  \bibinfo{year}{2021}\natexlab{}.
\newblock \showarticletitle{Modular, compositional, and executable formal
  semantics for {LLVM} {IR}}.
\newblock  \bibinfo{number}{{ICFP}} (\bibinfo{year}{2021}).
\newblock
\urldef\tempurl%
\url{https://doi.org/10.1145/3473572}
\showDOI{\tempurl}


\bibitem[\protect\citeauthoryear{Zhao, Nagarakatte, Martin, and Zdancewic}{Zhao
  et~al\mbox{.}}{2012}]%
        {vellvm}
\bibfield{author}{\bibinfo{person}{Jianzhou Zhao}, \bibinfo{person}{Santosh
  Nagarakatte}, \bibinfo{person}{Milo M.~K. Martin}, {and}
  \bibinfo{person}{Steve Zdancewic}.} \bibinfo{year}{2012}\natexlab{}.
\newblock \showarticletitle{Formalizing the {LLVM} intermediate representation
  for verified program transformations}. In
  \bibinfo{booktitle}{\emph{Proceedings of the Symposium on Principles of
  Programming Languages, {POPL}}}.
\newblock
\urldef\tempurl%
\url{https://doi.org/10.1145/2103656.2103709}
\showDOI{\tempurl}


\bibitem[\protect\citeauthoryear{Zhao, Nagarakatte, Martin, and Zdancewic}{Zhao
  et~al\mbox{.}}{2013}]%
        {ZhaoNMZ13}
\bibfield{author}{\bibinfo{person}{Jianzhou Zhao}, \bibinfo{person}{Santosh
  Nagarakatte}, \bibinfo{person}{Milo M.~K. Martin}, {and}
  \bibinfo{person}{Steve Zdancewic}.} \bibinfo{year}{2013}\natexlab{}.
\newblock \showarticletitle{Formal verification of SSA-based optimizations for
  {LLVM}}. In \bibinfo{booktitle}{\emph{{ACM} {SIGPLAN} Conference on
  Programming Language Design and Implementation, {PLDI} 2013}}.
  \bibinfo{publisher}{{ACM}}.
\newblock
\urldef\tempurl%
\url{https://doi.org/10.1145/2491956.2462164}
\showDOI{\tempurl}


\end{thebibliography}

\end{document}